\documentclass[epj]{svjour}
\usepackage{latexsym}
\usepackage{amssymb}
\usepackage{graphics}
\usepackage{url}
\usepackage{color}
\begin{document}
%
\title{Particle correlations from the initial state}
\author{Tolga Altinoluk\inst{1} \and N\'estor Armesto\inst{2}
}                     
\offprints{}          
\institute{National Centre for Nuclear Research, 02-093, Warsaw, Poland  \and Instituto Galego de F\'{\i}sica de Altas Enerx\'{\i}as IGFAE, Universidade de Santiago de Compostela, 15782 Santiago de Compostela, Galicia-Spain}
\date{Received: date / Revised version: date}
%
\abstract{
The observation in small size collision systems, $pp$ and $p$A, of strong correlations with long range in rapidity and a characteristic structure in azimuth, the ridge phenomenon, is one of the most interesting results obtained at the Large Hadron Collider. Earlier observations of these correlations in heavy ion collisions at the Relativistic Heavy Ion Collider are standardly attributed to collective flow due to strong final state interactions, described in the framework of viscous relativistic hydrodynamics. Even though  data for small size systems are well described in this framework, the applicability of hydrodynamics is less well grounded and initial state based mechanisms have been suggested to explain the ridge. In this review, we discuss particle correlations from the initial state point of view, with focus on the most recent  theoretical developments.
\PACS{
      {PACS-key}{discribing text of that key}   \and
      {PACS-key}{discribing text of that key}
     } 
} 
\maketitle
%

\section{Introduction}
\label{intro}
\begin{figure}
\resizebox{0.5\textwidth}{!}{%
  \includegraphics{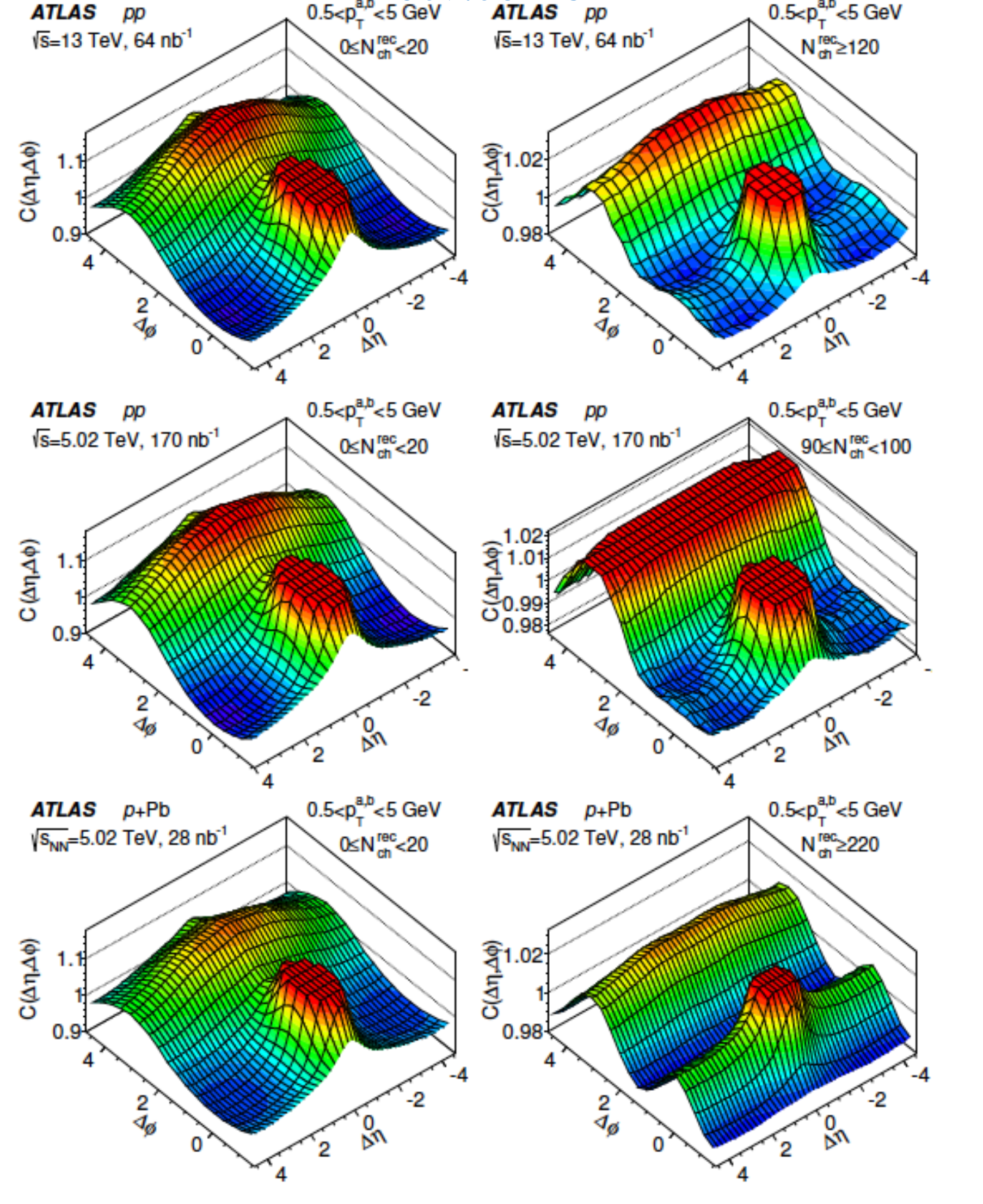}
}
\caption{Two particle correlations in $pp$ and $p$Pb collisions at the LHC measured by the ATLAS Collaboration~\cite{Aaboud:2016yar}, for different energies and particle multiplicities in the event. Taken from~\cite{Aaboud:2016yar}.}
\label{fig:1}       
\end{figure}

While the focus of the physics programme at the Large Hadron Collider (LHC) is the discovery and understanding of the properties of the previously missing piece in the Standard Model -- the Higgs boson -- and the search for its eventual failure, it has also shown very surprising and unexpected aspects of Quantum Chromodynamics (QCD), particularly in small collisions systems, $pp$ and $p$A. One of the most exciting observations made in high multiplicity $pp$ collisions by the CMS collaboration during the first LHC run is the discovery of the correlations between produced particles over large intervals of rapidity, peaking at zero relative azimuthal angle~\cite{Khachatryan:2010gv}. This phenomenon was dubbed {\it ridge€} due its shape in the azimuthal angle-rapidity plot, and constitute one of the key findings at the LHC (see Fig.~\ref{fig:1}). 

Later on, this structure was found by other collaborations and for smaller multiplicities~\cite{Aaboud:2016yar,Khachatryan:2015lva,Aad:2015gqa,Khachatryan:2016txc} and in association with $Z$ boson production~\cite{Aaboud:2019mcw}. A similar ridge structure was also observed in $p$Pb collisions at the LHC by the four large collaborations~\cite{CMS:2012qk,Abelev:2012ola,Aad:2012gla,Aaij:2015qcq}. A maximum in the correlations also appears at azimuthal angle $\pi$, called the away side ridge in contrast to the near side ridge peaked at zero azimuthal angle. They have also been observed in PbPb collisions, see e.g.~\cite{Chatrchyan:2013nka,Abelev:2014mda,Aaboud:2017acw} for PbPb results and a comparison with those in $p$Pb. Similar correlations were observed in AuAu, $d$Au and $^3$HeAu collisions at the Relativistic Heavy Ion Collider (RHIC)~\cite{Alver:2009id,Abelev:2009af,Adare:2014keg,Adamczyk:2015xjc,Adare:2015ctn,PHENIX:2018lia}. They have also been observed in photoproduction on Pb in ultraperipheral collisions (UPCs) at the LHC~\cite{ATLAS:2019gsn}. Their existence in smaller systems like $e^+e^-$ collisions~\cite{Badea:2019vey} at the Large Electron-Positron collider and deep inelastic scattering (DIS) events in $ep$ at the Hadron-Elektron-Ringanlage~\cite{ZEUS:2019jya} has been scrutinised, but the results are not conclusive. The ridge is the most striking feature in the long list of similarities between small and large collision systems in the observed results for many observables~\cite{Schlichting:2016sqo,Loizides:2016tew,Schenke:2017bog,Nagle:2018nvi,Citron:2018lsq}.

The standard explanation for such azimuthal asymmetries in heavy ion collisions (HICs) is the existence of strong final state interactions that lead to a situation where viscous relativistic hydrodynamics can be applied, see the reviews~\cite{Jeon:2016uym,Romatschke:2017ejr}. The dynamics leading to such a situation, called hydrodynamisation, is unclear~\cite{Romatschke:2016hle} and both strong and weak coupling explanations have been proposed, see e.g.~\cite{Keegan:2015avk} and refs. therein. Furthermore, the hydrodynamic description seems to hold for large anisotro\-pies, i.e. rather far from local equilibrium. 

The hydrodynamic description of the azimuthal asymmetries in $pp$ and $p$Pb collisions at the LHC is successful~\cite{Romatschke:2017ejr,Schenke:2012wb}. There is a hot ongoing discussion on the explanation of such success and of the fact that hydrodynamics seems to be the effective description for long wavelength modes in any field theory, see e.g.~\cite{Kurkela:2019set} and refs. therein. But it demands a very careful choice of initial conditions, specifically that the proton is modelled as a collection of constituent quarks or hot spots. The description seems to be pushed to the limit of small collision areas and low particle densities where non-hydrodynamic modes play a very important role, as seen in both hydrodynamic studies~\cite{Romatschke:2017ejr} and in those that consider a weak coupling quasiparticle picture in transport frameworks~\cite{Kurkela:2019kip}.

Therefore, the hydrodynamic explanation for the azimuthal asymmetries in small systems looks tenuous. Besides, causality arguments show that long range correlations in rapidity must come from the very early stages of the collision~\cite{Dumitru:2008wn}. And hydrodynamic calculations demand initial conditions that contains long range rapidity correlations, initial energy and particle density and flow profiles and, unless they are assumed to be completely washed out by final state interactions, correlations. So, beyond addressing the obvious fundamental question: \textit{is the strong interaction dynamics capable to lead to collectivity through final state interactions even in small systems, or is the origin of the ridge correlations different in $pp$ and $p$Pb collisions than in HICs?}, searching for correlations coming from the initial state is linked to the understanding of the dynamics prior to the use of hydrodynamics and to the provision of well grounded initial conditions for hydrodynamic calculations.

This contribution is devoted to the description and discussion of those frameworks which lead to correlations among partons in the initial stage that, if not washed out by  strong final state interactions and hadronisation -- that we will assume in the following, may lead to azimuthal asymmetries as observed in data. We start by those based on the weak coupling but non-perturbative description of dense partonic systems offered by the the Color Glass Condensate (CGC) effective theory, see the reviews~\cite{Iancu:2002xk,McLerran:2008uj,Gelis:2010nm} and the book~\cite{Kovchegov:2012mbw}. This will be the subject of Sections~\ref{sec:1} and \ref{sec:2}. We will then review other explanations inspired in QCD in Section~\ref{sec:3}, to end with a summary and discussions in Section~\ref{sec:4}.

Our focus will be on recent formal developments and we will base the presentation in our own works and formalism, trying to make connection with the other formalisms which differ in notation. We will comment briefly on the status of the comparison to experimental data in the summary.

\section{Two particle correlations from the CGC}
\label{sec:1}

The observation of the ridge correlations in small size systems has triggered a lot of efforts to understand whether the structure of the initial state itself can lead, in $pp$ and $p$A collisions, to such correlations without resourcing to final state interactions. Over the last decade, several mechanisms have been suggested to explain the ridge correlations in the CGC framework. The two most successful ones are the "domain structure of the target" developed in~\cite{Kovner:2012jm,Kovner:2010xk,Kovner:2011pe} and the "glasma graph approach" introduced in~\cite{Dumitru:2008wn,Armesto:2006bv,Dumitru:2010iy}\footnote{Apart from these two approaches, there are also other CGC-based mechanisms to describe the two particle correlations from the initial state. In~\cite{Levin:2011fb,Iancu:2017fzn}, it is argued that long range rapidity correlations can be explained by the spatial variation of the partonic density in the target. On the other hand, in~\cite{Abramovsky:2005vm}, the origin of  two particle correlations is explained through the rapidity evolution of  dipole operators by breaking the mean field approximation. Correlations in the hadron wave functions as described in the CGC have been recently considered in~\cite{Albacete:2018bbv} but with the aim of providing initial conditions for hydrodynamic evolution beyond simple energy, flow and particle density transverse profiles.}. 

The underlying mechanism for the domain structure of the target can be summarised as follows: the hadronic target is assumed to contain domains of oriented chromoelectric fields in the transverse plane. When two partons (normally assumed to be gluons when the scattering takes place at high energies and the probed values of momentum fraction of the partons, $x$, is small) from the projectile are close enough to scatter on the same domain, they get a common final momentum that reflects the correlated structure of the fields in the target. As gluons belong to the adjoint, thus real, representation of the $SU(N_c)$ group, the correlation holds for both parallel and antiparallel momenta, thus justifying the near and away side structures. The size of the domain in the target is assumed to be of order $1/Q_s$, with $Q_s$ being the saturation momentum which is the characteristic transverse momentum for the partons in the saturated target wave function described by the CGC~\cite{Iancu:2002xk,McLerran:2008uj,Gelis:2010nm}. Projectile partons lying closer than $1/Q_s$ contribute mainly to particle production in the region of transverse momentum $p_T\gtrsim Q_s$. Therefore, this mechanism should mainly be applicable in that transverse momentum region.

Note that this model implies a non-trivial target structure that goes beyond the usual isotropic averages employed in CGC calculations, see below. While still lacking justification from first principles (although indeed CGC numerical calculations indicate that field correlations in the hadron wave functions are characterised by length scales $\sim 1/Q_s$ \cite{Krasnitz:1999wc,Krasnitz:2000gz,Lappi:2003bi}), this explanation is often used for qualitative discussion and understanding of numerical results, and may have further implications on e.g. spin or Transverse Momentum Distributions (TMDs) physics. Numerical studies based on models containing this domain structure have been performed in \cite{Dumitru:2014vka,Dumitru:2014yza,Dumitru:2015cfa}. They show correlations that go beyond leading number of colours, see the discussion below, and lead to odd harmonics, see Section~\ref{sec:2}.

On the other hand, the glasma graph approach to two particle correlations is very successful to describe many features of the data as shown in \cite{Dusling:2012iga,Dusling:2012cg,Dusling:2012wy,Dusling:2013qoz,Dusling:2017dqg,Dusling:2017aot}, but the physics behind this approach was not clear. This issue has been studied in~\cite{Altinoluk:2015uaa} and it has been shown that a genuine quantum effect, Bose enhancement of the gluons in the projectile wave function, leads to final state correlations in the glasma graph approach\footnote{In~\cite{Blok:2017pui,Blok:2018xes} a collective behaviour and azimuthal asymmetries are obtained for small systems in a perturbative framework as a consequence  of quantum interference and colour correlations. Spatial anisotropies result into momentum anisotropies via multipole radiation patterns. This approach, albeit formulated in a rather different language, shows similarities with the glasma graph approach.}.

The concept of Bose enhancement  for a generic quantum system can be understood by considering a state with fixed occupation number, $\{n_i(p)\}$, of $N$ species of bosons at different momenta which, up to some normalisation factor, can be written as  
\begin{equation}
\big| \{ n_i(p)\}\big\rangle \propto \prod_{i,p}\big[ a_i^\dagger(p)\big]^{n_i(p)}|0\rangle,
\end{equation} 
with  $a^\dagger_i(p)$ the creation operator of the boson and $i=1,2,\dots, N$. The mean particle density $\tilde n$ is defined as the expectation value of the number operator in this state: 
 \begin{equation}
 \tilde n\equiv \big\langle \{ n_i(p)\} \big|\sum_j a^\dagger_j(x)a_j(x)\big| \{ n_i(p)\}\big\rangle=\sum_{i,p} n_i(p).
 \end{equation}
 The two particle correlator in momentum space $C(p,k)$ is defined in a similar way and can be calculated in a trivial manner:  
 \begin{eqnarray}
 \label{2_part_corr_Mom}
 C(p,k)&=&\Big[ \sum_in_i(p)\Big]\, \Big[ \sum_jn_j(k)\Big]\nonumber \\
 &+&\delta(p-k)\sum_i\big[n_i(p)\big]^2.
 \end{eqnarray}
The first term on the right hand side of Eq. (\ref{2_part_corr_Mom}) 
is the square of the mean particle density and the second term is the Bose enhancement term. It vanishes when the momenta of the two bosons are different and gives an enhancement when the momenta of two bosons coincide whi\-ch is ${\mathcal O}(1/N)$, due to the fact that it contains a single sum over the species index. The physics behind this is the fact that only bosons of the same species are correlated with each other. 
 
\begin{figure*}[htb]
\resizebox{1\textwidth}{!}{%
  \includegraphics{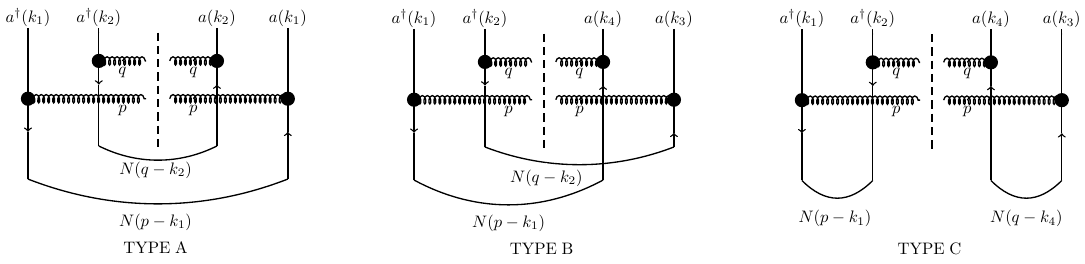}
\vspace*{3cm} 
}
\caption{Glasma graphs for two gluon inclusive production before averaging over the projectile colour charge density $\rho$. Black blobs denote vertices  and dashed lines the cuts separating the amplitudes (to the left of the cut) from the complex conjugate amplitudes (to the right). Taken from~\cite{Altinoluk:2015uaa}.}
\label{glasma}       
\end{figure*}

Let us now describe how Bose enhancement arises in the CGC and leads to final state correlations by considering the double inclusive gluon production within the glasma graph approach. In this approach each gluon is assumed to come from a different colour charge density in the projectile wave function that is rapidity invariant\footnote{These two assumptions are justified at high enough energy or at small $x$ where the colour charge density is high so gluons can be treated semiclassically, and they have evolved radiation tails that populate phase space uniformly in $\ln 1/x$.}. For our purposes, these colour charge densities can be conveniently represented in terms of gluon creation and annihilation operators in the incoming projectile wave function. After averaging over the target fields the glasma graphs can be written as sum of three types of diagrams (see Fig. \ref{glasma}). 

Type A diagrams describe the case when  two gluons with transverse momenta ${\bf k}_1$ and ${\bf k}_2$ scatter independently on the target, acquiring transfer of  transverse momentum ${\bf p-k}_1$ and ${\bf q-k}_2$ so that the outgoing gluons have transverse momenta ${\bf p}$ and ${\bf q}$. Type B and Type C diagrams include interference contributions which are also interesting to study but, for now, let us focus on the Type A contribution and show how the Bose enhancement effect can be observed by studying these diagrams alone. The Type A contribution to the double inclusive gluon production can be written as 
\begin{eqnarray}
\label{TypeA}
\!\!{\rm Type\,  A} \!&=& \!\!\int \frac{d^2{\bf k}_1}{(2\pi)^2}\frac{d^2{\bf k}_2}{(2\pi)^2} \, 
\langle {\rm in}| a^{\dagger i}_a({\bf k}_1) a^{\dagger j}_b({\bf k}_2) a^{k}_a({\bf k}_1)a^{l}_b({\bf k}_2)|{\rm in}\rangle \nonumber\\
&\times& \bigg[ \delta^{ik}-\frac{{\bf k}_1^i{\bf k}_1^k}{{\bf p}^2}\bigg]\, \bigg[ \delta^{jl}-\frac{{\bf k}_2^j{\bf k}_2^l}{{\bf q}^2}\bigg]
 N({\bf p}-{\bf k}_1)\, N({\bf q}-{\bf k}_2),\nonumber \\
\end{eqnarray}
where $|{\rm in}\rangle$ is the wave function of the incoming projectile and $N({\bf p-k})$ is the dipole scattering amplitude -- the scattering amplitude for a two gluon system to scatter on the target.  Moreover, the rapidity dependence of the gluon creation and annihilation operators is integrated over. The explicit dependence on rapidity becomes important only when the rapidity difference between the observed particles is parametrically large, $\Delta\eta\gtrsim 1/\alpha_s$.

The evaluation of the expectation value of any operator in the incoming projectile state requires a two averaging procedure in the CGC. 
In~\cite{Altinoluk:2015uaa},  averaging over the valence colour charge density is performed first which leads to the density matrix operator $\hat{\rho}$ on the soft gluon Hilbert space. Then, the second averaging over the soft gluons is performed using this density matrix operator. The two particle correlator that appears in the Type A contribution calculated with this procedure leads to the following result:
\begin{eqnarray}
\label{correlator}
&&
\hspace{-0.5cm}
C({\bf k}_1,{\bf k}_2)= S_{\perp}^2(N_c^2-1)^2 
 \frac{{\bf k}_1^i{\bf k}_1^k}{{\bf k}_1^2}
  \frac{{\bf k}_2^j{\bf k}_2^l}{{\bf k}_2^2}
  \frac{g^4\mu^2({\bf k}_1)\mu^2({\bf k}_2)}{{\bf k}_1^2{\bf k}_2^2} \nonumber\\
&&
\hspace{-0.5cm}
\times 
\bigg\{1+\frac{1}{S_\perp (N_c^2-1)}\Big[ \delta^{(2)}({\bf k_1-k_2})+\delta^{(2)}({\bf k_1+k_2})\Big]\bigg\},
\end{eqnarray} 
where $S_\perp$ is the transverse area of the projectile. The first term on the right hand side of Eq.~(\ref{correlator}) 
is the classical term which corresponds to the square of the number of gluons, while the second term is the typical Bose enhancement term, relatively suppressed by the number of states in the adjoint colour representation.

If we consider a situation where the incoming projectile has intrinsic saturation momentum $Q_s$ and the momenta of the produced gluons are also $\sim Q_s$, i.e. $|{\bf p}|\sim|{\bf q}|\sim Q_s$,  then the production amplitude is dominated by the contributions $|{\bf k}_1|\sim|{\bf k}_2|\sim Q_s$. The initial state correlations are encoded in the Bose enhancement terms in Eq.~(\ref{correlator}), 
which are delta functions. The interaction with the target is obtained by convoluting the two particle correlator with the dipole amplitudes $N({\bf p-k_1})N({\bf q-k_2})$. Since, in this kinematics, the momentum transfers from the target ( $|{\bf p-k_1}|\sim|{\bf q-k_2}|\ll Q_s$ ) are small and the Bose enhancement terms involve delta functions, these initial state correlations naturally transform into angular correlations between the produced gluons in the final state. In a more general case, the delta functions, which are an artefact of considering a translationally invariant projectile, are smeared when convoluted with the dipole scattering amplitudes but this should not completely destroy the final state angular correlations. 

The immediate question that arises after the study of gluons is whether quarks are subject to correlations in the CGC. This question has been posed in~\cite{Altinoluk:2016vax} where the correlations between the produced quarks were studied. The results in~\cite{Altinoluk:2015uaa} show that the origin of the correlations between the produced gluons is the Bose enhancement of the projectile gluons. Due to their fermionic nature, one expects quarks to experience Pauli blocking which effectively amounts to a suppression of the probability of finding two quarks with the same quantum numbers in the CGC state. Therefore, one should expect a negative correlation between the final state quarks that originate from the initial state ones.  On the other hand, the correlation between the gluons is found to be long range in rapidity since  the CGC wave function is dominated by the rapidity integrated soft gluon field. Thus, another important question to answer is:  are the (anti)correlations between the final state quarks  long or short range in rapidity?  The answer to this question is not obvious a priori. In the projectile wave function, quarks are produced via splitting of the rapidity  invariant gluons into quark-antiquark pairs. However, the splitting amplitude itself depends on the rapidity of the quark and antiquark. Moreover, due to this splitting in the projectile wave function the expression for the production cross section of quarks is much more complicated  compared to the one for gluons. These questions are answered in~\cite{Altinoluk:2016vax} where it was shown that the initial state correlations between the quarks in the projectile wave function are not distorted by the small momentum transfer from the target in specific kinematics. In these kinematics, the rapidity difference between the produced quarks is relatively large, i.e. $\eta_1-\eta_2\gg1$. Moreover, a large contribution comes from the situation where the transverse momenta of the produced quarks ${\bf p}$ and ${\bf q}$ are of the same order and much larger than the saturation scale of the projectile $Q_s$, and the saturation scale of the projectile is much larger than saturation scale of the target $Q_T$, i.e. $|{\bf p}|\sim |{\bf q}|\gg Q_s\gg Q_T$. Then the contribution to the production cross section that is sensitive to correlations has the following behaviour (see Eqs.~(3.19) and (3.20) in~\cite{Altinoluk:2016vax} for the full expressions):
\begin{equation}
\frac{d\sigma}{d^2{\bf p}d\eta_1\, d^2{\bf q}d\eta_2}\bigg|_{\rm corr.}\propto -\Big[e^{-(\eta_1-\eta_2)}(\eta_1-\eta_2)^2\Big].
\end{equation}
The negative sign of this contribution shows that it suppresses the classical term as opposed to the gluon case. This is the result of the Pauli blocking effect in quark-quark production. Moreover, this effect decays exponentially with the rapidity difference between the two produced quarks, which shows that it is short range in rapidity.  However, this exponential decrease is tempered by two powers of the rapidity difference.

\begin{figure}[h!]
\resizebox{0.5\textwidth}{!}{%
  \includegraphics{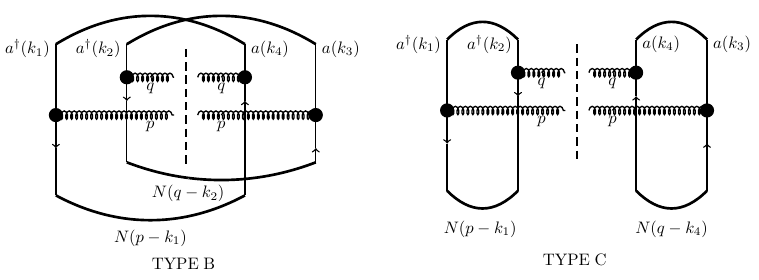}
}
\caption{Glasma graph diagrams (after averaging over the projectile colour charges) that lead to HBT correlations. Taken from~\cite{Altinoluk:2015eka}.}
\label{TypeBC}       
\end{figure}

Besides, there is another physical effect present in the glasma graph approach which is referred to as the Hanbury-Brown-Twiss (HBT) correlations between the produced gluons\footnote{HBT correlations are studied in~\cite{Kovchegov:2012nd,Kovchegov:2013ewa} in the $k_T$ factorisation approach.}~\cite{Altinoluk:2015eka}. The diagrams in the glasma graph approach that lead to HBT correlations are those in Fig.~\ref{TypeBC} after performing pair wise contraction of the colour charges in the projectile wave function.
Assuming a translationally invariant projectile wave function, the contribution from Type B and Type C diagrams to the production cross section is
\begin{equation}
{\rm Type\, B}\propto \delta^{(2)}({\bf p-q}) \; , \hspace{1cm} {\rm Type\, C}\propto \delta^{(2)}({\bf p+q}).
\end{equation}

If the translational invariance condition is relaxed, then the delta functions are smeared  
over a scale of the inverse size $R^{-1}$ of the projectile: 
$|{\bf p\pm q}|\sim R^{-1}$. This size $R$ represents the radius of the gluon cloud inside the proton and its inverse is smaller than the saturation scale, $R^{-1}< Q_s$.  Moreover, it is also shown in \cite{Altinoluk:2015eka} that the HBT correlations are long range in rapidity just as the Bose enhancement effect. Thus, the strength of the HBT correlations is equal when the rapidities of the two produced gluons are similar ($\eta_1\simeq \eta_2$) or when the difference between them is large ($|\eta_1-\eta_2|\gg1$).

\begin{figure}
\resizebox{0.5\textwidth}{!}{%
  \includegraphics{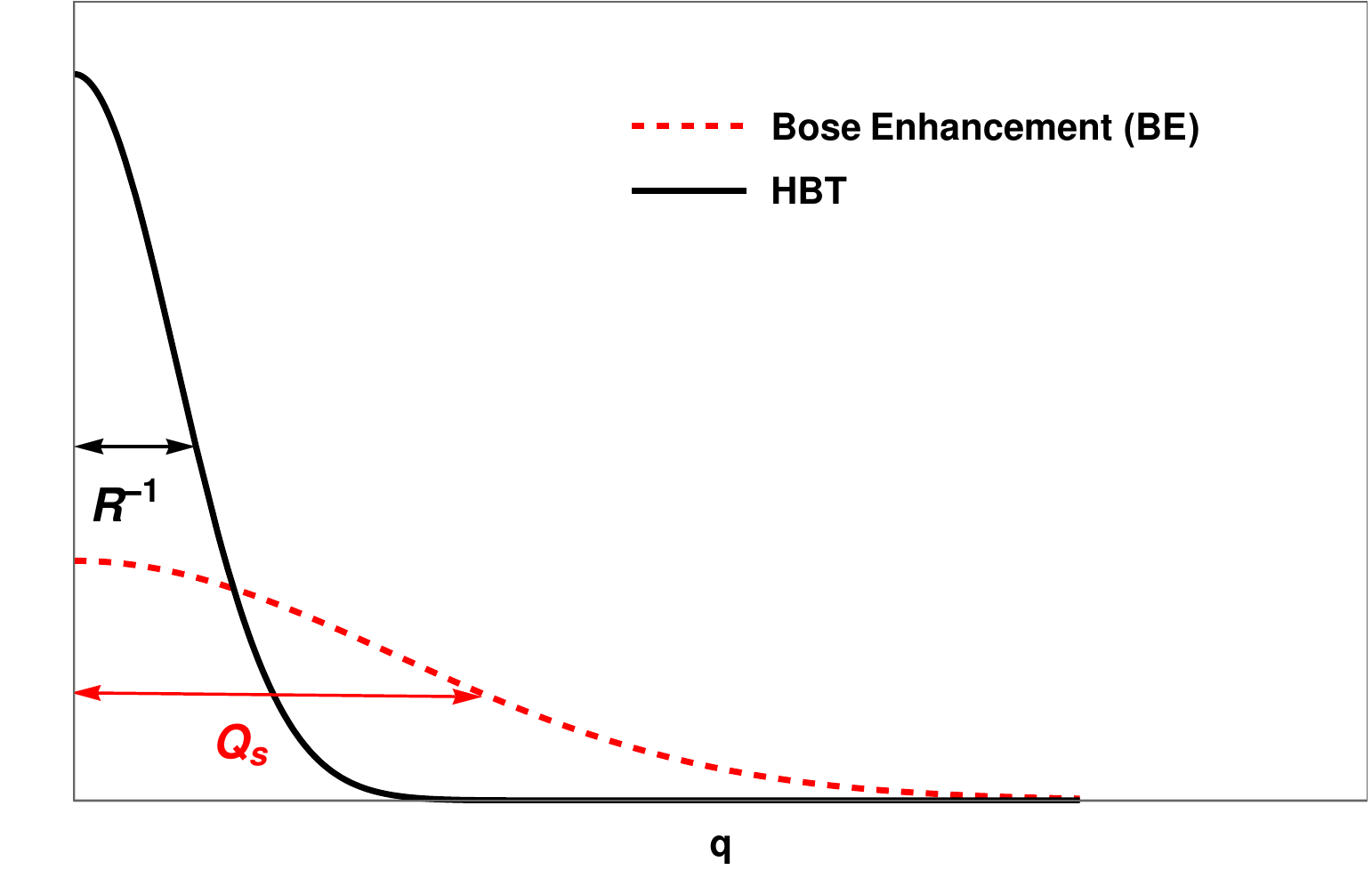}
}
\caption{Schematic separation in $q$ (here the modulus of the difference in transverse momentum between the produced gluons) between the contributions to the HBT effect (solid line) and  to the Bose enhancement effect (dashed line) in the two particle correlation function. Taken from~\cite{Altinoluk:2015eka}.}
\label{contr}       
\end{figure}

To sum up, the correlation function $C({\bf p,q})$, formally defined as the ratio of double inclusive gluon production cross section to the square of the single one, in the glasma graph approach contains two physical effects and can be written as follows: 
\begin{equation}
\label{Correlation}
C({\bf p,q})=1+C({\bf p,q})\Big|_{\rm BE}+C({\bf p,q})\Big|_{\rm HBT}\,.
\end{equation}
The first term on the right hand side of Eq.~(\ref{Correlation}) 
is the classical contribution which originates from the square of  single inclusive production. $C({\bf p,q})\big|_{\rm BE}$ represents the effect of  Bose enhancement of the gluons in the projectile wave function. As described above, this effect leads to a correlation of the final state gluons. On the other hand, $C({\bf p,q})\big|_{\rm HBT}$ represents the HBT correlations in the glasma graph approach which directly introduces correlations between the final state gluons. Both $C({\bf p,q})\big|_{\rm BE}$ and $C({\bf p,q})\big|_{\rm HBT}$ are rapidity independent, therefore long range in rapidity. The Bose enhancement contribution is suppressed by the transverse area of the projectile with respect to the HBT contribution (actually by the number of ``sources" $Q_s^2 S_\perp$). However, it leads to correlations whose width in momentum space is determined by the saturation momentum $Q_s$. On the other hand, the HBT contribution is not suppressed but it gives a narrow peak in momentum space with width $R^{-1}$.
This comparison is shown in Fig.~\ref{contr}. 

In the explicit calculations in the glasma graph approach to double inclusive particle production\footnote{Three and four gluon inclusive production are considered in \cite{Ozonder:2014sra,Ozonder:2017wmh} within the glasma graph approach.}, the averaging over the target configurations that leads to the dipole scattering amplitude is performed expanding this amplitude to the lowest non trivial order in the target field strength, corresponding to two gluon exchange between the gluons in the projectile wave function and the target. The dipole scattering amplitudes $N({\bf p-k_1})$ and $N({\bf q-k_2})$  introduced in Eq.~(\ref{TypeA})  
are assumed to originate from single pairs of target fields and therefore this approach does not take into account the effects of multiple scatterings in a dense target. Therefore, this approach is only valid for $pp$ collisions\footnote{The  previously discussed conditions in the projectile of large colour charge and rapidity independence of the gluon distribution are assumed to hold in $pp$.}. In~\cite{Altinoluk:2018ogz}, the inclusive production of two  and three gluons is computed beyond the glasma graph approach by including the multiple scattering effects,  which extends the validity of the glasma graph approach from $pp$ to $p$A collisions\footnote{This extension is studied numerically in~\cite{Lappi:2015vta}, and also analytically in~\cite{Altinoluk:2018hcu}. The main difference between the latter and~\cite{Altinoluk:2018ogz} is the computation framework. Two gluon correlations in~\cite{Altinoluk:2018hcu} are computed within the $k_T$-factorised approach which is difficult to generalise to three or more particles. Moreover, the results of~\cite{Altinoluk:2018hcu} are valid only in the large $N_c$ limit as opposed to the results in~\cite{Altinoluk:2018ogz} which are valid for finite $N_c$. In this sense,~\cite{Altinoluk:2018hcu} can be considered as the first attempt to generalise the glasma graph approach to two gluon production from $pp$ to $p$A.}.

Apart from taking into account the multiple scattering effects in~\cite{Altinoluk:2018ogz}, a systematic way to identify each term in the double inclusive gluon production cross section and to characterise whether it is a Bose enhancement or HBT contribution is introduced. This identification is performed by adopting the following strategy. When calculating the double inclusive gluon production, one has to average over four colour charges (two in the amplitude and two in the complex conjugate amplitude) in the projectile wave function: $\langle \rho^{a_1}({\bf x_1})\rho^{a_2}({\bf x_2})\rho^{b_1}({\bf y_1})\rho^{b_2}({\bf y_2})\rangle_P$. Here,  $({\bf x_i},a_i)$ and $({\bf y_i},b_i)$ stand for the transverse position and colour indices of the colour charge densities in the amplitude and in the complex conjugate amplitude, respectively.  The averaging over the colour charge distributions in the projectile is commonly performed by using a generalised McLerran-Venugopalan (MV) model~\cite{McLerran:1993ni,McLerran:1994vd} where the weight functional is Gaussian. Then, the average of any product of colour charge densities factorises into a product of all possible pair, Wick-like contractions. The correlator of two colour charge densities in momentum space can be defined as
\begin{equation}
\label{rhorho}
\langle \rho^a({\bf k})\rho^b({\bf p})\rangle_P=\delta^{ab}\mu^2({\bf k, p}).
\end{equation}
The function $\mu^2({\bf k, p})$ characterises the structure of the projectile. It can be written as 
\begin{equation}
\label{mu}
\mu^2({\bf k, p})=T\bigg( \frac{\bf k-p}{2}\bigg) F\big[ ({\bf k+p})R\big],
\end{equation}
where  $F\big[ ({\bf k+p})R\big]$ is a soft form factor with maximal value $F(0)$, and $R$ is the radius of the projectile. Function $T$ defines the transverse momentum dependent distribution of the valence charges. The soft form factor identifies whether a term is a contribution to the Bose enhancement of the projectile gluons or a contribution to the HBT correlations of the produced gluons. For example, in our set up the produced gluons have momenta ${\bf p}$ and ${\bf q}$, while the projectile gluons carry transverse momenta ${\bf k_1}$ and ${\bf k_2}$. In this case, $\mu^2({\bf p,q})$ gives a maximal contribution when ${\bf p+q}=0$ which clearly can be identified as the HBT correlations of the produced gluons. $\mu^2({\bf k_1,k_2})$ is peaked when ${\bf k_1+k_2}=0$  which is a contribution to the Bose enhancement of the gluons in the projectile wave function.  

On the other hand, multiple scattering effects on the dense target are taken into account by introducing the standard Wilson lines in the CGC fra\-mework. In this fra\-mework, the interaction between the projectile and the target is assumed to be eikonal which amounts to the situation where each parton produced by the projectile colour charge scatters on the target by picking up a colour rotation described by a Wilson line which is defined as an exponential of the target field ordered in the $x^+$ coordinate:
\begin{equation}
\label{Wilson_line}
U_{\cal R}({\bf x})={\cal P}_+\,  e^{ig\int dx^+\, T^a_{\cal R}\, A^-_a(x^+,{\bf x})},
\end{equation}
at the amplitude level. Here, $T^a_{\cal R}$ is the $SU(N_c)$ generator in the representation ${\cal R}$ which can be the fundamental one for a quark and the adjoint one for a gluon. This leads to the appearance in the cross section of  double dipole and quadrupole amplitudes (in the adjoint representation) of the type  
\begin{equation}
\langle s({\bf x,y})s({\bf z,w})\rangle_T\,,\hspace{1cm}  \langle Q({\bf x,y,z,w})\rangle_T \,,
\end{equation}
 which have to be averaged over the target field distributions. The dipole and the quadrupole operators are defined as 
\begin{eqnarray}
s({\bf x},{\bf y})&=&\frac{1}{N_c^2-1} {\rm tr}\big[U({\bf x})U^\dagger({\bf y})\big],\\
Q({\bf x,y,z,w})&=&\frac{1}{N_c^2-1}{\rm tr}\big[ U({\bf x})U^\dagger({\bf y})U({\bf z})U^\dagger({\bf w})\big].
\end{eqnarray} 

The cross section has to be integrated over four transverse coordinates. In principle, the maximal contribution should come from the area in coordinate space, i.e.  when all the four coordinates are far away from each other. However, all four points cannot be far away from each other since the target field ensemble has to be colour neutral, and colour neutralisation in the CGC happens on scales of order $1/Q_s$.
Therefore, the maximal contribution to the integral must come from the configurations where the four points are combined into pairs, such that each pair is a singlet and the distance between the pairs is large. This is the leading contribution in $1/(Q_s^2 R^2)$ to the integral on transverse coordinates\footnote{An equivalent reasoning based on colour neutralisation at large distances can be found in~\cite{Kovchegov:2012nd,Kovchegov:2013ewa}.} -- not, by any means, a good representation of the target averages of ensembles of Wilson lines by themselves. Taking into account only such configurations is equivalent to calculating the target averages of products of any number of Wilson lines by factorising them into averages of pairs with basic Wick contraction. In this case, target averaging of the double dipole and the quadrupole amplitudes can be written as 
\begin{eqnarray}
\label{factorised}
\langle Q({\bf x,y,z,w})\rangle_T&\approx& d({\bf x,y})d({\bf z,w}) +  d({\bf x,w})d({\bf z,y})\nonumber\\
&+&\frac{1}{N_c^2-1}d({\bf x,z})d({\bf y,w}),\\
\langle s({\bf x,y})s({\bf z,w})\rangle_T&\approx&d({\bf x,y})d({\bf z,w}) \nonumber\\
&&
\hspace{-2.5cm}
+\, 
\frac{1}{(N_c^2-1)^2}\Big[ d({\bf x,w})d({\bf z,y})+d({\bf x,z})d({\bf y,w})\Big],
\end{eqnarray}
where we have defined $d({\bf x,y})\equiv \langle s({\bf x,y})\rangle_T$. 
Then, by using the function $\mu^2({\bf k,p})$ given in Eq.~(\ref{mu}) 
for the projectile colour charge density correlators and using the factorisation ansatz described above for the double dipole and quadrupole amplitudes, the double inclusive gluon production cross section is computed and the nature of all the terms is identified. Moreover, it is also shown that the contributions to final state correlations comes from the quadrupole terms in two gluon production\footnote{A comprehensive study of gluon-gluon, quark-quark and quark-antiquark correlations in $p$A collisions is also performed in~\cite{Martinez:2018ygo,Martinez:2018tuf}.}.

\section{Odd azimuthal harmonics from the CGC}
\label{sec:2}

\begin{figure}[h!]
\resizebox{0.5\textwidth}{!}{%
  \includegraphics{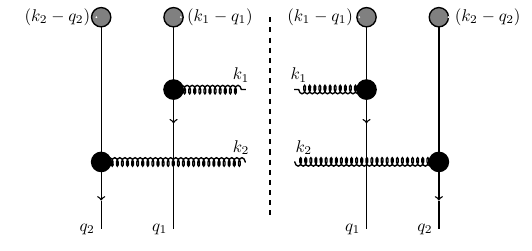}
}
\caption{Momentum assignment for the double inclusive gluon production. The grey blobs represent the colour charge densities in the projectile wave function. ${\bf q}_1$ and ${\bf q}_2$ are the transverse momenta transferred  from the target during the interaction.}
\label{double_inc}       
\end{figure}

To describe one key existing problem in usual CGC calculations, namely the absence of odd harmonics, let us briefly discuss double inclusive particle production in  more depth.
Within the approximations described in Section~\ref{sec:1}, double inclusive gluon production is computed in $p$A collisions in~\cite{Altinoluk:2018ogz}. The production cross section of two gluons with rapidities $\eta_1$ and $\eta_2$, and transverse momenta ${\bf k}_1$ and ${\bf k}_2$ (see Fig.~\ref{double_inc}) reads 
\begin{eqnarray}
\label{double_inc_full}
&&
\frac{d\sigma}{d^2{\bf k}_1d\eta_1 d^2{\bf k}_2d\eta_2}=\alpha_s^2(4\pi)^2(N_c^2-1)^2
\!\!\int \frac{d^2{\bf q}_1}{(2\pi)^2} \frac{d^2{\bf q}_2}{(2\pi)^2} d({\bf q}_1)
\nonumber\\
&&
\times \, 
d({\bf q}_2)
\bigg\{ I_0+\frac{1}{N_c^2-1}I_1+\frac{1}{(N_c^2-1)^2}I_2\bigg\} + ({\bf k}_2\to-{\bf k}_2),\nonumber \\ 
\end{eqnarray}
where 
\begin{eqnarray}
I_0&=&\frac{1}{2}\mu^2({\bf k}_1-{\bf q}_1, {\bf q}_1-{\bf k}_1) \, \mu^2({\bf k}_2-{\bf q}_2, {\bf q}_2-{\bf k}_2)
\nonumber\\
&
\times&\,  L^i({\bf k}_1, {\bf q}_1)L^i({\bf k}_1, {\bf q}_1) \, L^j({\bf k}_2, {\bf q}_2)L^j({\bf k}_2, {\bf q}_2),
\\
I_1&=& \mu^2\ ({\bf k}_1-{\bf q}_1, {\bf q}_2-{\bf k}_2) \, \mu^2({\bf k}_2-{\bf q}_2, {\bf q}_1-{\bf k}_1)\nonumber\\
&
\times&\, L^i({\bf k}_1, {\bf q}_1)L^i({\bf k}_1, {\bf q}_1) \, L^j({\bf k}_2, {\bf q}_2)L^j({\bf k}_2, {\bf q}_2)
\nonumber\\
&
+&\, 
\mu^2\ ({\bf k}_1-{\bf q}_1, {\bf q}_1-{\bf k}_2) \, \mu^2({\bf k}_2-{\bf q}_2, {\bf q}_2-{\bf k}_1)\nonumber\\
&
\times&\, L^i({\bf k}_1, {\bf q}_1)L^i({\bf k}_1, {\bf q}_2) \, L^j({\bf k}_2, {\bf q}_1)L^j({\bf k}_2, {\bf q}_2),
\\
I_2&=&\mu^2({\bf k}_1-{\bf q}_1, {\bf q}_2-{\bf k}_1) \, \mu^2({\bf k}_2-{\bf q}_2, {\bf q}_1-{\bf k}_2) 
\nonumber\\
&\times&\,  L^i({\bf k}_1, {\bf q}_1)L^i({\bf k}_1, {\bf q}_2) \, L^j({\bf k}_2, -{\bf q}_1)L^j({\bf k}_2, -{\bf q}_2) 
\nonumber\\
&+&\, \mu^2({\bf k}_1-{\bf q}_1, {\bf q}_2-{\bf k}_2) \, \mu^2(-{\bf k}_1-{\bf q}_2, {\bf q}_1+{\bf k}_2)
\\
&\times&\,  L^i({\bf k}_1, {\bf q}_1)L^i({\bf k}_1, -{\bf q}_2) \, L^j({\bf k}_2, -{\bf q}_1)L^j({\bf k}_2, {\bf q}_2).\nonumber
\end{eqnarray}
Here,  function $\mu^2$ defines the structure of the projectile, Eq.~(\ref{mu}). On the other hand,  function $L^i({\bf k},{\bf q})$ is the eik\-onal Lipatov vertex which is defined as 
\begin{equation}
\label{EikLipatov}
L^i({\bf k}, {\bf q})\equiv \frac{({\bf k}-\bf {q})^i}{({\bf k}-{\bf q})^2}-\frac{{\bf k}^i}{{\bf k}^2}\ .
\end{equation} 
Note that, as discussed in Section~\ref{sec:1}, all correlations are subleading in $1/N_c$ which is due to the use of Gaussian averages -- the MV model. Moreover, the double gluon inclusive production cross section is written in terms of the dipole averages assuming translational invariance of the target (a standard approximation which is reasonable in $p$A for large nuclei):
\begin{eqnarray}
d({\bf x}_1,{\bf x}_2)&=&\int \frac{d^2{\bf q}_1}{(2\pi)^2}\frac{d^2{\bf q}_2}{(2\pi)^2} e^{-i{\bf q}_1\cdot {\bf x}_1+i{\bf q}_2\cdot{\bf x}_2}\nonumber\\
&\times&
d\bigg(\frac{{\bf q}_1+{\bf q}_2}{2}\bigg)\delta^{(2)}({\bf q}_1- {\bf q}_2).
\end{eqnarray}

A convenient way to study two particle correlations is through a Fourier decomposition into harmonics defined in for the azimuthal angle $\Delta\phi$ between the produced gluons with transverse momenta ${\bf k}_1$ and ${\bf k}_2$. When Fourier expanded, the double inclusive gluon spectrum Eq.~(\ref{double_inc_full}) can be written as 
\begin{equation}
N(k_1,k_2, \Delta\phi)=a_0(k_1,k_2)\bigg[1+\sum_{n=0}^{\infty}2V_{n\Delta}(k_1,k_2)\cos(n\Delta\phi)\bigg],
\end{equation}
where 
\begin{equation}
V_{n\Delta}(k_1,k_2)=\frac{\int_0^{\pi} N(k_1,k_2,\Delta\phi)\cos(n\Delta\phi)d\Delta\phi}{\int_0^{\pi}N(k_1,k_2,\Delta\phi)d\Delta\phi}\ .
\end{equation}
One way of defining the $p_T$ dependence of the Fourier coefficients is by fixing one of the momenta ($k_1=p^{ref}_T$), and treating the other one as a free variable ($k_2=p_T$). With this choice, the azimuthal harmonics are defined as 
\begin{equation}
\label{def:v_n}
v_n(p_T)=\frac{V_{n\Delta}(p_t,p^{ref}_T)}{\sqrt{V_{n\Delta}(p^{ref}_t,p^{ref}_T)}}\ .
\end{equation}

The key theoretical problem for the description of the two particle correlations within the CGC is the {\it absence of the odd harmonics}. This problem is analyzed in~\cite{Kovner:2010xk}, and it is shown to be due to the symmetry $({\bf k}_2\to-{\bf k}_2)$ of the double inclusive gluon production cross section Eq.~(\ref{double_inc_full}) which is also referred to as the ``accidental symmetry of the CGC". 

Three ways\footnote{Here we refer to gluon production which is the dominant mechanism at small $x$ or high energies. Quarks, obeying Fermi-Dirac statistics and belonging to a non-real colour representation, can give rise to odd harmonics as investigated in~\cite{Dusling:2017aot,Davy:2018hsl,Zhang:2019dth}.} have been proposed to solve this problem\footnote{Besides, the role of the centrality or multiplicity event selection for the breaking of the accidental symmetry and the appearance of odd azimuthal harmonics has been analysed in~\cite{Gotsman:2018mde}.}. On the one hand, the projectile and target can be characterised by a more involved structure than that considered in the usual MV averages~\cite{Kovner:2012jm,Kovner:2010xk,Kovner:2011pe,Dumitru:2014vka,Dumitru:2014yza,Dumitru:2015cfa}.

On the other hand and as discussed previously, within the CGC framework each produced gluon originates from a separate colour charge density in the projectile wave function as shown in Fig.~\ref{double_inc}. The contributions to the projectile wave function that emerge from merging of the gluons before the interaction with the target or splitting of a gluon into two gluons emitted from the same colour charge density, are not taken into account in the standard CGC calculations. Recently, in~\cite{Kovner:2016jfp,Kovchegov:2018jun} it is shown that the accidental symmetry of the CGC can be broken by including such corrections to the projectile wave function. The even and odd parts of the double inclusive gluon production cross section under the accidental symmetry are computed separately, and finally, the azimuthal harmonics are calculated. The corresponding numerical studies and a comparison with data are performed in~\cite{Mace:2018yvl,Mace:2018vwq}. 

Finally, inclusive gluon production is usually studied within the eikonal approximation in the CGC framework. In recent studies~\cite{Agostini:2019avp,Agostini:2019hkj}, it is shown that the accidental symmetry can be broken by going beyond this eikonal approximation. In the next subsection, we introduce a systematic way to include subeikonal corrections in  CGC calculations and show how these corrections give rise to non-vanishing odd harmonics.

\subsection{Subeikonal corrections in the CGC}

In inclusive gluon production at central rapidity in $p$A collisions, both the projectile and the target are highly energetic since they are boosted from their initial rapidity to the central rapidity where the collision occurs. Therefore, in this case both colliding objects can be treated in the CGC framework. This corresponds to defining the projectile  by the colour charge $J^\mu_a(x)$,
\begin{equation}
\label{colour_charge_eik}
J^\mu_a(x)=\delta^{\mu+}\delta(x^-)\rho_a({\bf x}),
\end{equation}
and  the target by the colour field $A^\mu_a(x)$ that is given as 
\begin{equation}
\label{field_eik}
A^{\mu}_a(x)=\delta^{\mu-}\delta(x^+)A^-_a({\bf x}).
\end{equation} 
Let us recall that these expressions of the colour charge of the projectile and the colour field of the target are defined within the eikonal approximation, which is justified by the large energy of both colliding objects. If for the dilute projectile the eikonal approximation can be trusted at a given energy, the same approximation for a large target can be true only for larger energies. The eikonal approximation for the target amounts to the following three conditions: 
\begin{enumerate}
\item $A^{\mu}_a(x)\simeq \delta^{\mu -}A^-_a(x)$: Neglecting the (+) and transverse components of the colour field of the target.  
\item $A^{\mu}_a(x)\simeq A^{\mu}_a(x^+,{\bf x})$: Neglecting the $x^-$ dependence in the colour field of the target.
\item $A^\mu(x)\propto \delta(x^+)$: Assuming that the target field is peaked around  $x^+=0$ due to Lorentz contraction, which is also known as the shockwave approximation. 
\end{enumerate}
In realistic kinematical conditions under which the experiments are performed, the energies are not asymptotic and the eikonal approximation is not always justified. While for a dilute projectile it is usually valid even for high energy collisions,  this is not necessarily true for a large nucleus. Relaxing any of the above approximations accounts for corrections to the eikonal limit. In~\cite{Altinoluk:2014oxa,Altinoluk:2015gia}, a systematic method to compute the corrections to the eikonal limit by relaxing the third approximation is  developed. This corresponds to treating the colour field of the target with a finite longitudinal support $L^+$ along the $x^+$ direction, thus replacing Eq.~(\ref{field_eik}) by 
\begin{equation}
\label{field_noneik}
A^{\mu}_a(x)\simeq \delta^{\mu-}A^-_a(x^+,{\bf x}).
\end{equation}
Such subeikonal corrections are thus subleading with respect to the infinite Lorentz contraction of the target. 

Before discussing the results, let us give a brief sketch of the method employed to derive the non-eikonal corrections. Let us consider the production of a single gluon with transverse momenta ${\bf k}$ and longitudinal momenta $k^+$ in $p$A collisions at central rapidity. The dilute projectile is still treated in the eikonal approximation and defined with the charge density $J^\mu_a(x)$ given in Eq.~(\ref{colour_charge_eik}). On the other hand, the eikonal approximation is relaxed for the dense target that is defined by the colour field $A^\mu_a(x)$ given in Eq.~(\ref{field_noneik}) with a finite support from $0$ to $L^+$ in the longitudinal direction. In this case, the production cross section can be written as the square of the gluon production amplitude averaged over the projectile and target distributions and integrated over impact parameter ${\bf B}$:  
 \begin{equation}
 2k^+\frac{d\sigma}{dk^+d^2{\bf k}}= \int d^2{\bf B}\sum_{\lambda} \bigg\langle \Big\langle |{\cal M}^a_\lambda(\underline{k}, {\bf B})|^2\Big\rangle_P\bigg\rangle_T\ .
 \end{equation}   
Here, $\lambda$, $a$ and ${\underline k}=(k^+, {\bf k})$  are the polarization, colour and momentum of the produced gluon\footnote{Hereafter, we use the underline notation to indicate that for coordinates $\underline x=(x^+,{\bf x})$ and for momentum $\underline k=(k^+,{\bf k})$. }. For a target with finite longitudinal width, the gluon production amplitude ${\cal M}^a_\lambda(\underline{k}, {\bf B})$ is composed of three different contributions: gluon production before, while and after the projectile propagates through the target. At leading order, it is possible to relate the total gluon production amplitude and the background retarded gluon propagator by using the LSZ reduction formula and the perturbative expansion of the colour field of the target~\cite{MehtarTani:2006xq}. In the light cone gauge  $A^+=0$, the total gluon production amplitude can be written in terms of the $(i-)$ component of the background retarded gluon propagator $G_R^{\mu\nu}(x,y)$ as 
\begin{eqnarray}
\label{gluon_prod_amplitude}
{\cal M}^a_{\lambda}(\underline k,{\bf B})&=&\epsilon^{i*}_\lambda (2k^+)\lim_{x^+\to0}\int d^2{\bf x}\int dx^- e^{ik\cdot x}
\nonumber\\
&&\times\, 
\int d^4y\,  G^{i-}_R(x,y)_{ab} \, J^+_b(y).
\end{eqnarray}
Since the colour field of the target is independent of $x^-$, one can introduce the one-dimensional Fourier transform of the background retarded gluon propagator and write it in terms of of the background scalar propagator ${\cal G}_{k^+}^{\mu \nu}({\underline x}, {\underline y})$. Then, the $(i-)$ component of the retarded background gluon propagator reads
\begin{equation}
\label{retarded}
G_R^{i-}(x,y)_{ab}=\!\!\int \frac{dk^+}{2\pi}e^{-ik^+(x^- -y^-)}\frac{i}{2(k^++i\epsilon)^2}\partial_{{\bf y}^i}{\cal G}_{k^+}^{ab}(\underline x, \underline y).
\end{equation}
The background scalar propagator ${\cal G}_{k^+}^{ab}(\underline x, \underline y)$ satisfies the scalar Green's equation whose solution formally can be written as a path integral
\begin{eqnarray}
\label{scalar_prop}
{\cal G}_{k^+}^{ab}(\underline x, \underline y)&=&\theta(x^+-y^+)\int _{{\bf z}(y^+)={\bf y}}^{{\bf z}(x^+)={\bf x}}\big[ {\cal D}{\bf z}(z^+)\big] \, \\
&&\times\, 
e^{\frac{ik^+}{2}\int_{y^+}^{x^+} dz^+ {\bf \dot z}^2(z^+)} 
U^{ab}\Big( x^+,y^+;\big[{\bf z}(z^+)\big]\Big), \nonumber
\end{eqnarray}
with the Wilson line 
\begin{eqnarray}
&&U^{ab}\Big( x^+,y^+;\big[{\bf z}(z^+)\big]\Big)\nonumber\\
&&
={\cal P}_+\, \exp \bigg\{{ig\int_{y^+}^{x^+}\, d\tilde{ z}^+\, T^c\, A^-_c\Big( \tilde{z}^+,{\bf z}(z^+)\Big)}\bigg\}^{ab}
\end{eqnarray}
following the Brownian trajectory ${\bf z}(z^+)$. In the limit of vanishing longitudinal width,  $x^+-y^+\to 0$, the background scalar propagator ${\cal G}_{k^+}^{ab}(\underline x, \underline y)$ reduces to the standard Wilson line introduced in  Eq.~(\ref{Wilson_line}) 
and one recovers the eikonal limit. Therefore, it can be safely concluded that all the non-eikonal effects that are due to the finite longitudinal width of the target are encoded in the background scalar propagator. This also means that an expansion of ${\cal G}_{k^+}^{ab}(\underline x, \underline y)$ can be performed in terms of an eikonal parameter, with the first term in this expansion corresponding to the eikonal limit and  higher order terms to the corrections to this limit. 

 \begin{figure*}
\resizebox{1\textwidth}{!}{%
  \includegraphics{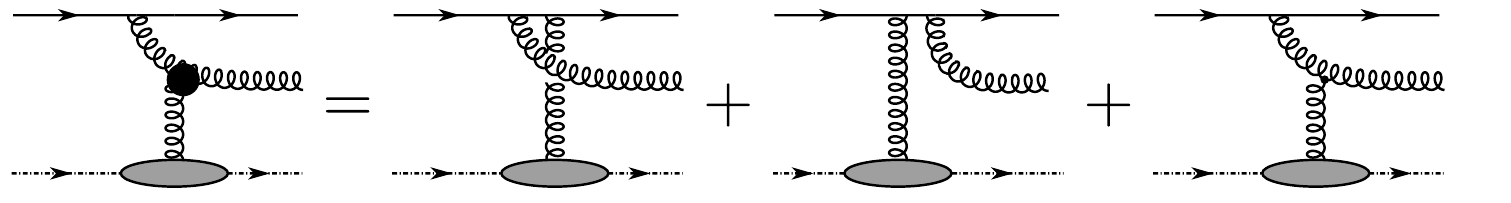}
\vspace*{3cm} 
}
\caption{Diagrams that contribute to the computation of the Lipatov vertex. The black blob represents the Lipatov vertex which is the sum of all real diagrams for gluon production shown on the right hand side of the equation. Taken from~\cite{Agostini:2019avp}.}
\label{mtotal}       
\end{figure*}

In order to perform an eikonal expansion of the background scalar propagator ${\cal G}_{k^+}^{ab}(\underline x, \underline y)$, one should first discretise the scalar background propagator. In the eikonal limit, $k^+/(x^+-y^+)$ is much larger than any transverse scale in the problem. In the large $k^+$ limit, it is natural to consider a generic path as a perturbation around the classical free path,
\begin{equation}
{\bf z}_n={\bf z}_n^{\rm cl}+{\bf u}_n\ ,
\end{equation}
where the transverse positions at step $n$ are on the straight line
\begin{equation}
{\bf z}_n^{\rm cl}={\bf y}+\frac{n}{N}({\bf x}-{\bf y})
\end{equation}
between the initial and final points, and the perturbation ${\bf u}_n$ satisfies the boundary conditions ${\bf u}_0={\bf u}_N=0$ with $N$ being the number of discretised steps. Once the expansion around the free classical path is performed for fixed initial and final positions, one should perform another expansion for small ${\bf x}-{\bf y}$, since ${\bf x}-{\bf y}$ is parametrically small in the large $k^+$ limit. 
After performing these two expansions up to second order in $(x^+-y^+)$ -- the finite longitudinal width of the target, the scalar background propagator ${\cal G}_{k^+}^{ab}(\underline x, \underline y)$ reads
\begin{eqnarray}
\label{expanded_Scalar}
&&
\int d^2{\bf x} \, e^{-i{\bf k}\cdot {\bf x}} \, {\cal G}_{k^+}^{ab}(\underline x, \underline y)
\nonumber\\
&&=\theta(x^+-y^+)e^{-i{\bf k}\cdot {\bf y}}e^{-k^-(x^+-y^+)}
\Big\{U(x^+,y^+; {\bf y})
\nonumber\\
&&
+\, \frac{(x^+-y^+)}{k^+}\Big[{\bf k}^i U^i_{[0,1]}(x^+,y^+; {\bf y}) + \frac{i}{2}U_{[1,0]}(x^+,y^+; {\bf y})\Big]
\nonumber\\
&&
+\,  \frac{(x^+-y^+)^2}{(k^+)^2} \Big[ {\bf k}^i{\bf k}^j U^{ij}_{[0,2]}(x^+,y^+; {\bf y}) \\
&&\hspace{1.0cm}
+\frac{i}{2}{\bf k}^iU^i_{[1,1]}(x^+,y^+; {\bf y})
-\frac{1}{4}U_{[2,0]}(x^+,y^+; {\bf y})\Big]\Big\}^{ab}.\nonumber
\end{eqnarray}
The first term on the right hand side of Eq.~(\ref{expanded_Scalar}) 
is the standard Wilson line defined in Eq.~(\ref{Wilson_line}).
${\cal O}\big[(x^+-y^+)/k^+\big]$ terms are the first order corrections to the strict eikonal limit which we refer to as next-to-eikonal (NEik) corrections. Similarly,  ${\cal O}\big[ (x^+-y^+)^2/(k^+)^2\big]$ terms are the second order corrections and they are referred to as next-to-next-to-eikonal (NNEik) corrections. The terms that are denoted as $U_{[\alpha,\beta]}(x^+,y^+; {\bf y})$ are the {\it decorated} Wilson lines which only appear beyond strict eikonal order. The first subscript $\alpha$ in the decorated Wilson lines stands for the order of expansion around the classical path while the second subscript $\beta$ stands for the order of the expansion around the initial transverse position ${\bf y}$.  The reason why these objects are referred to as {\it decorated} Wilson lines is related with their structure. These objects involve a background field insertion into the standard Wilson lines along the $+$-direction in a given $+$-coordinate. For example, the first decorated Wilson line is defined as 
\begin{eqnarray}
\label{decorated_Wilson}
&&
\hspace{-0.6cm}
\Big[U_{[0,1]}^{i}(x^+,y^+; {\bf y})\Big]^{ab}=\int_{y^+}^{x^+} dz^+ \frac{z^+-y^+}{x^+-y^+} 
\\
&&
\hspace{-0.6cm}
\times\, 
U^{ac}(x^+,z^+; {\bf y}) \big[ ig\, T^e_{cd}\, \partial_{{\bf y}^i}A^-_e(z^+,{\bf y})\big]\, U^{db}(z^+,y^+; {\bf y}).\nonumber
\, \end{eqnarray}  
The other decorated Wilson lines have similar structure with one or more background field insertions. We do not present the structure of all the decorated Wilson lines due their complexity and lengthy expressions (see~\cite{Altinoluk:2015gia}). One can easily get the expression for the gluon production amplitude at NNEik accuracy given in Eq.~(\ref{gluon_prod_amplitude}) 
by using the expression of the retarded background gluon propagator Eq.~(\ref{retarded}) 
and the expression derived for the background scalar propagator Eq.~(\ref{expanded_Scalar}). 

As discussed above, the retarded background gluon propagator $G_R^{\mu\nu}(x,y)_{ab}$ and, therefore, the scalar background propagator ${\cal G}_{k^+}^{ab}(\underline x, \underline y)$ are the main building blocks for computing cross sections in high energy $p$A collisions. In~\cite{Altinoluk:2014oxa,Altinoluk:2015gia}, these propagators were used to calculate the single inclusive gluon production cross section in $p$A collisions at NNEik accuracy. The same formalism can be adopted to compute double inclusive gluon production and hence the azimuthal harmonics in $p$A collisions~\cite{Agostini:2019avp,Agostini:2019hkj}.

In~\cite{Altinoluk:2015xuy}, the results of the single inclusive gluon production cross section at NNEik accuracy in $p$A collisions are used to study the weak field limit (i.e. glasma graph approximation) of this result which corresponds to single inclusive production in $pp$ collisions. In this limit, the decorated Wilson lines are expanded to first order in the background field of the target $A_a^-(z^+,{\bf y})$. For example, the first decorated Wilson line given in Eq.~(\ref{decorated_Wilson}) reduces to 
\begin{eqnarray}
\hspace{-1.2cm}
\Big[U_{[0,1]}^{i}(x^+,y^+; {\bf y})\Big]^{ab}&&\to \int_{y^+}^{x^+} dz^+\, \frac{z^+-y^+}{x^+-y^+}\, 
\nonumber\\
&&
\hspace{0.8cm}
\times\, 
\big[ ig\, T^c_{ab}\,  \partial_{{\bf y}^i}A^-_c(z^+,{\bf y})\big].
\end{eqnarray}
This simplification allows us to calculate the Lipatov vertex at NNEik accuracy. After expanding the eikonal and non-eikonal terms to first order in powers of the background field, which corresponds to the glasma graph approach in usual CGC calculations, the Lipatov vertex at NNEik accuracy can be written as 
\begin{eqnarray}
\label{NonEik_Lip}
L^i_{\rm NNEik}({\underline k},{\bf q};x^+)&=&
\bigg[\frac{({\bf k}-\bf {q})^i}{({\bf k}-{\bf q})^2}-\frac{{\bf k}^i}{{\bf k}^2}\bigg] \nonumber\\
&&
\hspace{-0.7cm}
\times\, 
\bigg\{1+i\frac{{\bf k}^2}{2k^+}x^+-\frac{1}{2}\bigg(\frac{{\bf k}^2}{2k^+}x^+\bigg)^2\bigg\}.
\end{eqnarray}
The first term on the right hand side of Eq.~(\ref{NonEik_Lip}) 
corresponds to the strict eikonal limit, thus it gives the eikonal Lipatov vertex defined in Eq.~(\ref{EikLipatov}). The second and the third terms are the NEik and NNEik corrections respectively. The structure of the vertex suggests that the corrections to the amplitude due to finite width of the target may exponentiate. 

This observation is further studied recently in~\cite{Agostini:2019avp} whe\-re it was shown that indeed the non-eikonal corrections due to finite longitudinal width of the target in the weak field limit exponentiate and can be written as modified Lipatov vertex. By computing the corresponding three diagrams (see Fig.~\ref{mtotal}) at the amplitude level and keeping the phase $e^{ik^-x^+}$ which is taken to be one in the eikonal limit, the non-eikonal Lipatov vertex that accounts for all order corrections to the eikonal limit due to finite longitudinal width of the target in the weak field limit reads  
\begin{equation}
L^i_{\rm NonEik}({\underline k}, {\bf q};x^+)=\bigg[\frac{({\bf k}-\bf {q})^i}{({\bf k}-{\bf q})^2}-\frac{{\bf k}^i}{{\bf k}^2}\bigg] 
e^{ik^-x^+}\ ,
\end{equation}
where $k^-\equiv {\bf k}^2/(2k^+)$. This structure was observed in the context of jet quenching in \cite{Wiedemann:2000za,Gyulassy:2000er,MehtarTani:2011gf} previously, however the identification of this building block for its use to include non-eikonal corrections in CGC calculations is done in~\cite{Agostini:2019avp} for the first time, further illustrating the close relation between CGC and jet quenching calculations~\cite{MehtarTani:2006xq}.

It is now straightforward to compute the non-eikonal single inclusive gluon production in $pp$ (i.e. dilute-dilute) collisions which formally reads
\begin{eqnarray}
&&
\frac{d\sigma}{d^2{\bf k}d\eta}\bigg|_{\rm dilute}^{\rm NonEik}= 4\pi \alpha_s\, C_A \, g^2\int dx_1^+dx_2^+ 
\int \frac{d^2{\bf q}_1}{(2\pi)^2} \frac{d^2{\bf q}_2}{(2\pi)^2} \nonumber\\
&&\times\, 
\delta^{c\bar c}\big\langle A^-_c(x_1^+, {\bf q}_1)A^-_{\bar c}(x_2^+, {\bf q}_2) \big\rangle_T \; 
\mu^2\big[{\bf k}-{\bf q}_1, {\bf q}_2-{\bf k}\big] 
\nonumber\\
&&
\times \, 
L^i_{\rm NonEik}({\underline k}, {\bf q}_1;x_1^+)\, L^i_{\rm NonEik}({\underline k}, {\bf q}_2;x_2^+).
\end{eqnarray}
An additional modification that is needed to account for the finite longitudinal width of the target is to adopt a modified expression for the correlator of two target fields. Motivated by the non zero longitudinal extent of the target, the fields can be located at different positions that are separated by the colour correlation length in the target $\lambda^+$. In this case, the two target field correlator reads
\begin{eqnarray}
\label{modfiedcorrelator}
&&
\hspace{-0.9cm}
\langle A^-_c(x_1^+, {\bf q}_1)A^-_{\bar c}(x_2^+, {\bf q}_2) \big\rangle_T=\delta^{c\bar c}\, n(x_1^+)\, \frac{1}{2\lambda^+}
\\
&&\times \, 
\theta\big(\lambda^+-|x_1^+-x_2^+|\big) \, (2\pi)^2 \delta^{(2)}({\bf q}_1-{\bf q}_2)\, |a({\bf q}_1)|^2\ , \nonumber
\end{eqnarray}
where  function $n(x^+)$ defines the one-dimensional target density that we take as constant, $n_0=n(x^+)$ for $0\leq x^+\leq L^+$, and $0$ elsewhere. Moreover,  function $a({\bf q})$ is the potential in momentum space which can be taken to be of Yukawa type, i.e. $|a({\bf q})|^2=m^2/({\bf q}^2+m^2)^2$ with $m$ being the Debye screening mass or inverse colour correlation length. In the eikonal limit, when $\lambda^+\to0$ for a constant potential and constant one-dimensional target density, one recovers the standard MV expression for the two target field correlator.  Using Eq.~(\ref{modfiedcorrelator}) one can integrate over the longitudinal coordinates that appear in the phases in non-eikonal Lipatov vertices. The final result of the non-eikonal single inclusive gluon production cross section in $pp$ collisions then reads
\begin{eqnarray}
\label{NonEikSingle}
&&
\hspace{-0.8cm}
\frac{d\sigma}{d^2{\bf k}d\eta}\bigg|_{\rm dilute}^{\rm NonEik}= 4\pi \alpha_s\, C_A (N_c^2-1)\, g^2 {\cal G}_1^{\rm NE}(k^-;\lambda^+)
\\
&&
\hspace{-0.5cm}
\times
\int\frac{d^2{\bf q}}{(2\pi)^2} \,  \mu^2\big[{\bf k}-{\bf q}, {\bf q}-{\bf k}\big] \, 
L^i({\bf k},{\bf q}) L^i({\bf k},{\bf q})\, 
|a({\bf q})|^2\ ,\nonumber
\end{eqnarray}
where we assume that the longitudinal width of the target is much larger than the colour correlation length, $\lambda^+\ll L^+$. In the cross section, Eq.~(\ref{NonEikSingle}), all the non-eikonal effects are encoded in the function ${\cal G}_1^{\rm NE}(k^-;\lambda^+)$ which is defined as 
\begin{equation}
\label{eq:G1}
{\cal G}_1^{\rm NE}(k^-;\lambda^+)=\frac{1}{k^-\lambda^+}\sin(k^-\lambda^+),
\end{equation}
with $k^-\equiv{\bf k}^2/2k^+$. In the limit of vanishing $(k^-\lambda^+)$ we have 
\begin{equation}
\lim_{k^-\lambda^+\to0}{\cal G}_1^{\rm NE}(k^-;\lambda^+)=1,
\end{equation}
and the well known eikonal limit for the single inclusive gluon production in the dilute target limit is recovered. Therefore, function ${\cal G}_1^{\rm NE}(k^-;\lambda^+)$ can be interpreted as the function that accounts for the relative importance of the non-eikonal effects with respect to the eikonal limit of the single inclusive gluon production in the dilute target limit.
\begin{figure}[h!]
\resizebox{0.45\textwidth}{!}{%
  \includegraphics{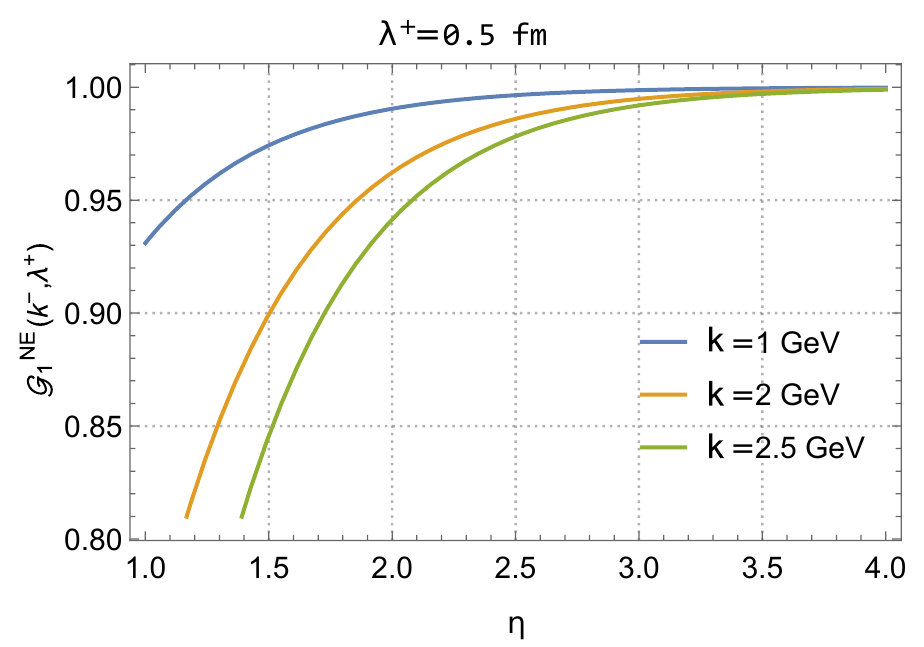}
}
\caption{The ratio of non-eikonal to eikonal single inclusive gluon production as a function of the rapidity of the produced gluon for different values of its transverse momenta at fixed correlation length $\lambda^+=0.5$ fm. Taken from~\cite{Agostini:2019avp}.}
\label{G1_noneik}       
\end{figure}

 In Fig. \ref{G1_noneik}, the ratio of the non-eikonal to eikonal single inclusive gluon production cross sections, i.e.  function ${\cal G}_1^{\rm NE}(k^-;\lambda^+)$, is plotted as a function of rapidity for different values of the transverse momenta of the produced gluon at correlation length $\lambda^+=0.5$ fm. The results show that with increasing rapidity of the produced gluon, the effects of the non-eikonal corrections vanish as expected from analytical predictions. Up to rapidity $\eta=2.5$, the relative importance of the corrections varies between 15\% and 2\% depending on the value of the transverse momenta.

Non-eikonal double inclusive gluon production cross section in $pp$ scattering can be computed in a similar manner. The main difference between the single and double inclusive production is that one needs to compute the target average of the four field correlator. This can be accomplished by factorising the average of the colour fields of the target into all possible Wick contractions which can be written
\begin{eqnarray}
&&
\big\langle A^-_a(x_1^+,{\bf q}_1)A^-_b(x_2^+,{\bf q}_2)A^-_c(x_3^+,{\bf q}_3)A^-_d(x_4^+,{\bf q}_4) \big\rangle_T
\\
&&
= \big\langle  A^-_a(x_1^+,{\bf q}_1)A^-_b(x_2^+,{\bf q}_2)\big\rangle_T
\big\langle A^-_c(x_3^+,{\bf q}_3)A^-_d(x_4^+,{\bf q}_4) \big\rangle_T
\nonumber\\
&&
+ \big\langle A^-_a(x_1^+,{\bf q}_1)A^-_d(x_4^+,{\bf q}_4) \big\rangle_T
\big\langle A^-_c(x_3^+,{\bf q}_3) A^-_b(x_2^+,{\bf q}_2) \big\rangle_T
\nonumber\\
&&
+\big\langle A^-_a(x_1^+,{\bf q}_1)A^-_c(x_3^+,{\bf q}_3)\big\rangle_T
\big\langle A^-_b(x_2^+,{\bf q}_2)A^-_d(x_4^+,{\bf q}_4) \big\rangle_T\,,
\nonumber
\end{eqnarray}
where each two target field correlator is defined in Eq.~(\ref{modfiedcorrelator}). The integrals over the longitudinal coordinates can be performed using this definition and the final result for the non-eikonal double inclusive gluon production cross section in $pp$ scattering can be organised in the following way:
\begin{eqnarray}
\label{NonEikDoubleInc}
&&
\frac{d\sigma}{d^2{\bf k}_1d\eta_1d^2{\bf k}_2d\eta_2}\bigg|_{\rm dilute}^{\rm NonEik}=\alpha_s^2 \, (4\pi)^2 \, g^4\, C_A^2\, (N_c^2-1)
\nonumber\\
&& \times
\int \frac{d^2{\bf q}_1}{(2\pi)^2} \frac{d^2{\bf q}_2}{(2\pi)^2} |a({\bf q}_1)|^2 |a({\bf q}_2)|^2 
{\cal G}_1^{\rm NE}(k_1^-;\lambda^+)
{\cal G}_1^{\rm NE}(k_2^-;\lambda^+) 
\nonumber\\
&&
\times
\bigg\{ I^{(0)}_{\rm 2tr}+ \frac{1}{N_c^2-1}\Big[I^{(1)}_{\rm 2tr}+I^{(1)}_{\rm 1tr}\Big]\bigg\},
\end{eqnarray}
where the subscripts denote the single trace terms ($I^{(i)}_{\rm 1tr}$) or the double trace term ($I^{(i)}_{\rm 2tr}$) which are analogue to double dipole and quadrupole operators in $p$A scattering discussed previously. The explicit expressions for these contributions read
\begin{eqnarray}
\label{I0}
I^{(0)}_{\rm 2tr}&=&
\mu^2\big[ {\bf k}_1-{\bf q}_1,{\bf q}_1-{\bf k}_1\big] 
\mu^2\big[ {\bf k}_2-{\bf q}_2,{\bf q}_2-{\bf k}_2\big] 
\nonumber\\
&
\times& \, 
L^i({\bf k}_1,{\bf q}_1)L^i({\bf k}_1,{\bf q}_1)L^j({\bf k}_2,{\bf q}_2)L^j({\bf k}_2,{\bf q}_2),
\\
\label{I1}
I^{(1)}_{\rm 2tr}&=&\Big\{ {\cal G}_2^{\rm NE}(k_1^-,k_2^-;L^+)
\mu^2\big[{\bf k}_1-{\bf q}_1, {\bf q}_2-{\bf k}_1\big]
\nonumber\\
&
\times&\, 
\mu^2\big[{\bf k}_1-{\bf q}_1, {\bf q}_2-{\bf k}_1\big]
L^i({\bf k}_1,{\bf q}_1)L^i({\bf k}_1,{\bf q}_2)
\nonumber\\
&
\times&\, 
L^j({\bf k}_2,{\bf q}_2)
L^j({\bf k}_2,{\bf q}_1)
\Big\} +({\underline k}_2\to-{\underline k}_2)
\end{eqnarray}
and, finally,
\begin{eqnarray}
\label{I2}
I^{(1)}_{\rm 1tr}&=& \Big\{ 
\mu^2\big[ {\bf k}_1-{\bf q}_1,{\bf q}_2-{\bf k}_2 \big]
\mu^2\big[ {\bf k}_2-{\bf q}_2,{\bf q}_1-{\bf k}_1 \big]  
\nonumber \\
&\times&\, 
L^i({\bf k}_1,{\bf q}_1)L^i({\bf k}_1,{\bf q}_1)L^j({\bf k}_2,{\bf q}_2)L^j({\bf k}_2,{\bf q}_2)
\nonumber\\
&+& {\cal G}_2^{\rm NE}(k_1^-,k_2^-;L^+) 
\Big\lgroup
\mu^2\big[ {\bf k}_1-{\bf q}_1,{\bf q}_1-{\bf k}_2 \big]
\nonumber\\
&\times&\, 
\mu^2\big[ {\bf k}_2-{\bf q}_2,{\bf q}_2-{\bf k}_1 \big]  
+\frac{1}{2} \mu^2\big[ {\bf k}_1-{\bf q}_1,{\bf q}_2-{\bf k}_2 \big]
\nonumber\\
&\times&\, 
\mu^2\big[ {\bf q}_2-{\bf k}_1,{\bf q}_1-{\bf k}_2 \big]  \Big\rgroup
L^i({\bf k}_1,{\bf q}_1)L^i({\bf k}_1,{\bf q}_2)
\nonumber\\
&\times&\, 
L^j({\bf k}_2,{\bf q}_1)L^j({\bf k}_2,{\bf q}_2)\Big\}+ ({\underline k}_2\to-{\underline k}_2).
\end{eqnarray}

In this setup (see Fig.~\ref{double_inc}), ${\bf k}_1-{\bf q}_1$ and ${\bf k}_2-{\bf q}_2$ are the transverse momenta of the two gluons in the projectile, ${\bf k}_1$ and ${\bf k}_2$ are the transverse momenta of the produced gluons in the final state, and ${\bf q}_1$ and ${\bf q}_2$ are the transverse momenta transferred from the target to the projectile during the interaction. By using the definition of function $\mu^2({\bf p},{\bf k})$ given in Eq.~(\ref{mu}) and the behaviour of the soft form factor, one can easily identify each term in the non-eikonal double inclusive  gluon production cross section given in Eqs.~(\ref{I0}), (\ref{I1}) and (\ref{I2}). Clearly, the term in $I^{(0)}_{\rm 2 tr}$ corresponds to the square of the single inclusive production and does not give any contribution to the correlations. The terms in $I^{(1)}_{\rm 2 tr}$ corresponds to the Bose enhancement of the target gluons since the soft form factor is peaked around ${\bf q}_1={\bf q}_2$. Finally, the terms in $I^{(1)}_{\rm 1 tr}$ contribute to the HBT correlations of the produced gluons and to Bose enhancement of the projectile gluons. 

In the non-eikonal double inclusive gluon production cross section, two functions appear that account for the non-eikonal effects: ${\cal G}_1^{\rm NE}(k_1^-;\lambda^+)$ presented in Eq.~(\ref{eq:G1}) and a new function ${\cal G}_2^{\rm NE}(k_1^-,k_2^-;L^+)$. This new function is defined as 
\begin{eqnarray}
\label{G2}
&&
\hspace{-1.5cm}
{\cal G}_2^{\rm NE}(k_1^-,k_2^-;L^+)
\nonumber\\
\hspace{1cm}
&&= \bigg\{ \frac{2}{(k_1^--k_2^-)L^+}\sin\bigg[ \frac{(k_1^--k_2^-)}{2}L^+\bigg]\bigg\}^2
\end{eqnarray}
which in the eikonal limit, i.e. $L^+\to 0$, goes to unity,
\begin{equation}
\lim_{L^+\to0}{\cal G}_2^{\rm NE}(k_1^-,k_2^-;L^+)=1.
\end{equation}
Different from the eikonal  double inclusive gluon production cross section, in the non-eikonal expression the mirror images are given by ${\underline k}_2\to-{\underline k}_2$ where ${\underline k}_2\equiv (k_2^+,{\bf k}_2)$. The mirror images of the terms that are accompanied by the function ${\cal G}_2^{\rm NE}(k_1^-,k_2^-;L^+)$ are now accompanied by ${\cal G}_2^{\rm NE}(k_1^-,-k_2^-;L^+)$. However, as obvious from the definition given in Eq.~(\ref{G2}), this function is not symmetric under this transformation. Moreover, in certain kinematic regimes the behaviour of ${\cal G}_2^{\rm NE}(k_1^-,k_2^-;L^+)$ differs completely from ${\cal G}_2^{\rm NE}(k_1^-,-k_2^-;L^+)$. Particularly, in a kinematic region where $k_1^-\sim k_2^-$, one gets 
\begin{equation}
{\cal G}_2^{\rm NE}(k_1^-,k_2^-;L^+)\gg {\cal G}_2^{\rm NE}(k_1^-,-k_2^-;L^+).
\end{equation}
This creates an asymmetry between the terms (${\underline k}_1,{\underline k}_2$) and their partners (${\underline k}_2\to-{\underline k}_2$). This asymmetry which comes from the non-eikonal effects reminds the asymmetry between the forward and backward peaks of the ridge structure observed in two particle production. Therefore, non-eikonal corrections break the accidental symmetry present in usual CGC calculations and can give rise to odd harmonics. In the remaining of the section we briefly examine the numerical relevance of this effect.

A detailed numerical analysis of the azimuthal structures in two particle correlations based on the non-eikonal double inclusive gluon production cross section given in Eq.~(\ref{NonEikDoubleInc}) with Eqs.~(\ref{I0}), (\ref{I1}) and (\ref{I2}), is performed in~\cite{Agostini:2019hkj} where it is assumed that:
\begin{enumerate}
\item the colour sources inside the projectile have a Gaussian distribution such that $\mu^2({\bf k}, {\bf q})=\mu^2(2\pi)^2\delta^{(2)}({\bf k}+{\bf q})$, with $\mu$ being the width of the Gaussian;
\item the Yukawa type potential that defines the target field correlators is given by $|a({\bf q})|^2=\mu_T^2/({\bf q}^2+\mu^2_T)^2$, with $\mu_T$ being an infrared regulator analogous to a Debye mass;
\item the transverse area of the projectile $S_\perp$ is defined thro\-ugh $(2\pi)^2\delta^{(2)}({\bf q}-{\bf q})\to S_{\perp}$. 
\end{enumerate}
\begin{figure}[h!]
\resizebox{0.5\textwidth}{!}{%
  \includegraphics{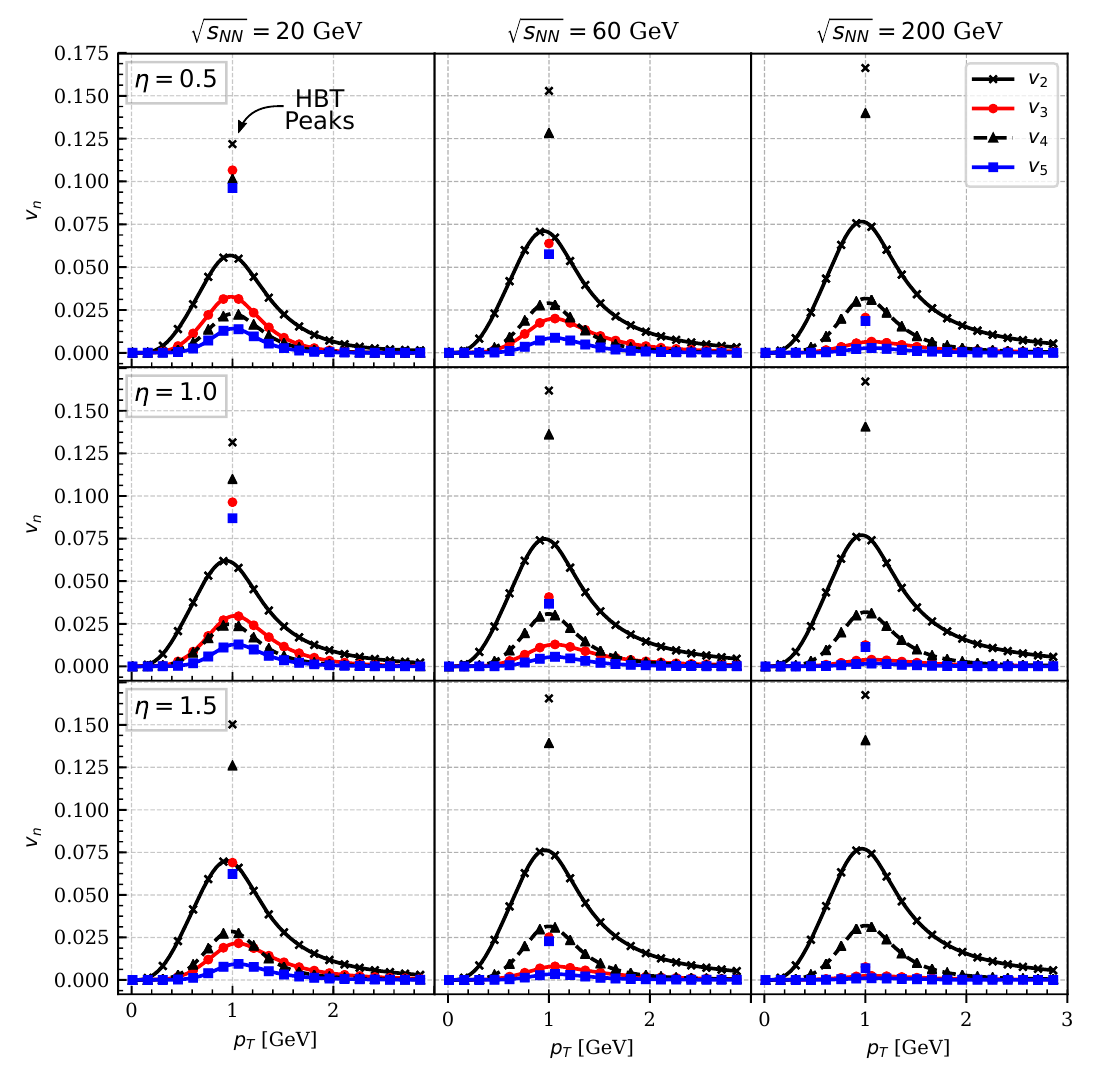}
}
\caption{Two particle azimuthal harmonics computed in the non-eikonal Glasma graph approach using the definition Eq.~(\ref{def:v_n}). The values are calculated  taking $\mu_T=0.4 \; {\rm GeV}$, $\mu_P=0.2 \; {\rm GeV}$ and $p_T^{ref}=1\; {\rm GeV}$ for different values of centre-of-mass energy and different gluon rapidities $\eta_1=\eta_2=\eta$. Taken from~\cite{Agostini:2019hkj}.}
\label{VnVsPt}       
\end{figure}
In this analysis,  function ${\cal G}_2^{\rm NE}(k_1^-,k^-_2;L^+)$ that encodes the non-eikonal effects defined in Eq.~(\ref{G2}) is rewritten as 
\begin{eqnarray}
&&
{\cal G}_2^{\rm NE}(k_1^-,k_2^-;L^+)
\\
&&= \bigg\{ \frac{\sqrt{2}}{(k_1e^{-\eta_1}-k_2e^{-\eta_2})L^+}
\sin\bigg[\frac{(k_1e^{-\eta_1}-k_2e^{-\eta_2})}{\sqrt{2}}L^+\bigg]\bigg\}^2 \nonumber
\end{eqnarray}  
 using $k^-=k^2/2k^+$, $k^+=ke^\eta/{\sqrt{2}}$, $k=|{\bf k}|$~\footnote{Assuming that $L$ is the size of the target in its rest frame, in the centre-of-mass frame $L^+$ is taken as 
\begin{equation}
L^+=\frac{1}{\gamma\sqrt 2}L\approx2A^{1/3}/\gamma\, {\rm fm} \approx 10 A^{1/3}/\gamma\, {\rm GeV^{-1}},
\end{equation}
where $A$ is the mass number of the nucleus and $\gamma\simeq\sqrt{s_{NN}}/(2m_N)$ accounts for the Lorentz contraction in the centre-of-mass frame. Moreover, for the numerics the gluonic size of the projectile is taken to be $B_p=4\, {\rm GeV^{-2}}$, the transverse size of the projectile is assumed to be $S_{\perp}=2\pi B_p\approx 9.8\,  {\rm mb}$ and the size of the target in its rest frame for a Pb nucleus is taken to be $L=12\, {\rm fm}$. Finally, the number of colours is taken to be $N_c=3$ and the colour correlation is set to be $\lambda^+=0$.}.

In Fig.~\ref{VnVsPt} the azimuthal harmonics up to $v_5$ are computed by using the definition given in Eq.~(\ref{def:v_n}). $p_T^{ref}$ is taken to be $1\, {\rm GeV}$ for different values of $\sqrt{s_{NN}}$ and $\eta_1=\eta_2=\eta$. 
The plot shows that the value of the odd harmonics decreases with increasing centre-of-mass energy at fixed rapidity $\eta$. This behaviour is the natural outcome of the fact that non-eikonal corrections become smaller with the increasing Lorentz gamma factor. Therefore, one can conclude that the non-eikonal corrections can be negligible for collisions at high centre-of-mass energy such as the ones at the LHC but they can be important for collisions at RHIC with $\sqrt{s_{NN}}\leq 200\, {\rm GeV}$. On the other hand, at any fixed energy the value of odd azimuthal harmonics decreases with increasing rapidity $\eta$. This behaviour is also expected, since the value of the odd harmonics is directly related to the non-eikonal corrections. The eikonal expansion parameter can be written as $p_TL^+e^{-\eta}$ in terms of  rapidity and, therefore, non-eikonal corrections (and thus the value of odd harmonics) decrease with increasing rapidity and vanish completely in the strict eikonal limit\footnote{In Fig. \ref{VnVsPt}, the unrealistic peaks that account for the HBT contributions are due to the use of $\mu^2({\bf k},{\bf q})\propto \delta^{(2)}({\bf k}+{\bf q})$. In a more realistic treatment, $\mu^2$ can be chosen as Gaussian which would peak around ${\bf k}+{\bf q}=0$ and show a bell shape behaviour.}.

\begin{figure}[h!]
\resizebox{0.5\textwidth}{!}{%
  \includegraphics{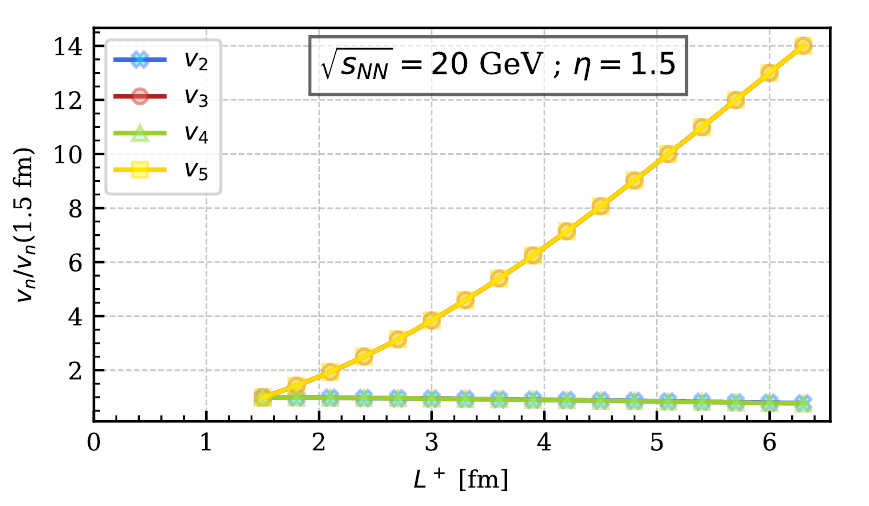}
}
\caption{Scaling of $v_n(L^+)/v_n(1.5\: {\rm fm})$ with $L^+$. Taken from~\cite{Agostini:2019hkj}.}
\label{RatioVsL}       
\end{figure}
 
In Fig.~\ref{RatioVsL}, the ratio $v_n(L^+)/v_n(1.5\: {\rm fm})$ is plotted as a function $L^+$ which reveals a very interesting feature of the effects of non-eikonal corrections on azimuthal harmonics: odd harmonics depend strongly on the size of the target while even ones are almost independent of it. Even though the explicit relation between multiplicity and $L^+$ requires a more dedicated study, the scaling of the odd harmonics with $L^+$ in Fig.~\ref{RatioVsL} qualitatively resembles the results of the analysis performed in~\cite{Mace:2018yvl} where it is shown that the value of $v_3$ increases with the increasing multiplicity. 
 
\section{Non-CGC explanations}
\label{sec:3}

Besides explanations to the ridge phenomenon based on the CGC, there are others that address its origin in the initial state of the collision or, at least, do not demand hydrodynamics or transport at work. It must be noted that the existence of long range rapidity correlations was discussed long ago as a consequence of multiple scattering, see~\cite{Capella:1978rg,Capella:1991mp}.

This approach was pushed forward in string models for multiparticle production, see e.g.~\cite{Amelin:1994mf}. Later on, several models that consider string interactions were argued to lead to azimuthal asymmetries: string percolation~\cite{Braun:2010tq,Andres:2014bia,Bautista:2009my} with the creation of azimuthally anisotropic strong chromoelectric fields, colour reconnection~\cite{Ortiz:2013yxa} that is able to produce some of the QGP-like features observed in $pp$, and string repulsion~\cite{Bierlich:2017vhg,Duncan:2019poz}. It is not clear whether the dynamics contained in these approaches can be considered as pure initial state but they offer a mechanism to produce the ridge in collisions between small systems that does not require any explicit final state rescattering, see~\cite{Nagle:2017sjv} for a model that explicitly requires parton and/or hadron rescattering to build  azimuthal asymmetries even with just two strings. In all these approaches, particle production from a single string is still isotropic and the anisotropy is built after string breaking.

A string-based model is also proposed in~\cite{Bjorken:2013boa,Glazek:2018gqs}. There, valence diquark-quark flux tubes or strings in the incoming protons overlap and produce more particles in the transverse than in the longitudinal  direction of the flux tube. Such anisotropic particle production leads to azimuthal asymmetries and the prediction has been made that it should also be visible in photoproduction, with particle production becoming maximal in the plane of the deflected electron (in $ep$ collisions) or proton (in UPCs).

It should be noted that all these approaches are inspired in the string behaviour of the QCD interaction in the non-perturbative domain, in contrast to the CGC that relies on perturbation theory for a small coupling constant. Indeed already in the framework of Reggeon field theory, some ideas have been pushed~\cite{Boreskov:2008uy} on the spatial variation of the transverse density in the hadron that resemble those in the CGC. Or CGC arguments have been extended to the soft physics domain and applied to describe azimuthal correlations, see~\cite{Gotsman:2016whc,Gotsman:2016fee} and subsequent papers of this group\footnote{There are also attempts to describe the near side ridge as a consequence of the momentum kick given by the leading parton to medium constituents~\cite{Wong:2011qr}, with a medium already present in $pp$ collisions, or to minijets~\cite{Trainor:2011cq}.}. 

Finally, let us indicate that azimuthal asymmetries arise in several processes when the nucleon is studied and characterised beyond collinear parton densities, as in the framework of Wigner distributions and TMD parton densities, see~\cite{Dumitru:2015gaa,Hatta:2016dxp}. Azimuthal asymmetries then arise in final observables like dijet production in DIS~\cite{Altinoluk:2015dpi,Dumitru:2016jku,Mantysaari:2019vmx}. But, although these calculations are often performed in a framework close to that of the CGC which is related with the TMD framework at small $x$~\cite{Altinoluk:2019wyu}, it goes beyond the standard CGC context to link with other physics like spin.

\section{Summary and discussions}
\label{sec:4}

In this manuscript we have discussed the explanations that are currently proposed to describe particle correlations, the ridge, observed in experimental data in small collision systems, $pp$ and $p$A, from the initial state point of view. Our main focus has been those weak coupling explanations based on the CGC. We have assumed that correlations among partons in the initial stage leave an imprint on those among particles in the final state, i.e. they are not washed out by final strong final state interactions and hadronisation.

First, we have focused on the standard eikonal treatment within the glasma graph approximation which is valid for collisions between dilute objects -- $pp$. We have reviewed  the studies which have shown that this approximation encodes two different type of contributions, namely the Bose enhancement of both projectile and target gluons and also HBT correlations of the produced final gluons. 

We have summarised the procedure that should be adopted to extend the validity of glasma graph approximation from dilute-dilute to dilute-dense (i.e. from $pp$ to $p$A) collisions by taking into account the multiple scattering effects in the dense target. We have shown that the structure of the double inclusive gluon production cross section is symmetric under (${\bf k}_2\to-{\bf k}_2$), which is known as the accidental symmetry of the CGC. Since this symmetry is the reason for vanishing odd harmonics in the CGC framework, we have discussed the suggested mechanisms to break this accidental symmetry.

In particular, we have focused on a specific mechanism to break this symmetry which is based on going beyond the standard eikonal approximation and including the subeikonal corrections that are due to the finite longitudinal width of the target. We have argued that such non-eikonal corrections, when included in the glasma graph approach to two particle correlations, successfully generate non-zero odd harmonics in specific kinematics. We would like to emphasise here that we make no attempt to compare the results with experimental data but only address the existence and  size of the non-eikonal effects on the azimuthal structure. As expected from non-eikonal corrections, their value and thus that of the odd harmonics decrease rapidly with increasing centre-of-mass energy. This decrease is strong since the analysis is performed for a dilute target -- a slower decrease of the size of the odd harmonics with increasing energy could be expected in a dilute-dense collision. Besides, the treatment of such non-leading eikonal corrections shows explicitly the link of the formalisms used in CGC and jet quenching calculations.

At this point, we should comment briefly on the comparison with experimental data. The main characteristics concerning azimuthal asymmetries in small systems observed in experiment~\cite{Khachatryan:2010gv,Aaboud:2016yar,Khachatryan:2015lva,Aad:2015gqa,Khachatryan:2016txc,CMS:2012qk,Abelev:2012ola,Aad:2012gla,Aaij:2015qcq,Aaboud:2017acw,Adare:2014keg,Adamczyk:2015xjc,Adare:2015ctn,PHENIX:2018lia} are:
\begin{itemize}
\item The  even and odd harmonics extracted using correlations between two and more particles, are of similar size to those found in larger systems.
\item They show the same dependence on the mass of the measured hadron as found in larger systems (see~\cite{Schenke:2016lrs} for an approach in the CGC).
\item Even harmonics show a much weaker dependence on the multiplicity in the event than odd harmonics.
\item $v_2$ and $v_3$ found in $p$Au, $d$Au and $^3$HeAu collisions at RHIC show, for central (head-on) collisions, the ordering $v_2^{p\mathrm{Au}}<v_2^{d\mathrm{Au}}\approx v_2^{^3\mathrm{HeAu}}$, $v_3^{p\mathrm{Au}}\approx v_3^{d\mathrm{Au}}< v_3^{^3\mathrm{HeAu}}$.
\item Measurements of many particle cumulants show evidence of collectivity. For example, four particle cumulants $c_2\{4\}=\langle e^{i2(\phi_1+\phi_2-\phi_3-\phi_4)}\rangle-2\langle e^{i2(\phi_1-\phi_2)} \rangle ^2$ ($v_2\{4\}=[-c_2\{4\}]^{1/4}$)
change sign from positive to negative with increasing associated multiplicity, with a smooth behaviour from small to large systems and from smaller to larger energies.
\end{itemize}

In the glasma graph approximation, several studies were done that describe $pp$ data in a reasonable manner~\cite{Dusling:2012iga,Dusling:2012cg,Dusling:2012wy,Dusling:2013qoz}. Later on, these studies were extended to $p$A with diverse degree of modelling \cite{Dusling:2017dqg,Dusling:2017aot}. Then, odd azimuthal harmonics were introduced following~\cite{Kovner:2016jfp,Kovchegov:2018jun}, with the corresponding numerical studies and a comparison with data performed in~\cite{Mace:2018yvl,Mace:2018vwq}. In these latter studies a successful comparison with RHIC and LHC data was initially claimed, which was later corrected after the criticism in~\cite{Nagle:2018ybc}. Nevertheless, it must be stated that none of the numerical calculations can be considered as a full implementation of the theoretical framework and that some results are still to be clarified from an analytical point of view, e.g. those in~\cite{Dusling:2017dqg,Dusling:2017aot} about the second Fourier coefficient defined through four particle correlations ($v_2\{4\}$) where the mentioned change of sign in $c_2\{4\}$ is attributed to multiple scattering beyond glasma graphs.

Finally, we have also shortly commented on approaches that are not based on, or go beyond, CGC ideas, to study the two particle correlations from the initial state. 

To conclude, let us indicate that future experimental programmes and facilities~\cite{RHICII,Citron:2018lsq,Accardi:2012qut,AbelleiraFernandez:2012cc,Benedikt:2018csr} will address the physics of small systems and the transition from small to large, particularly the onset and understanding of collectivity which is a central question in QCD at high energies.

\section*{Acknowledgements}
We would like to Ian Balitsky, Boris Blok, Giovannni Chirilli, Adrian Dumitru, Cyrille Marquet, Mauricio Martinez, Larry McLerran, Alfred Mueller, Jamie Nagle, Carlos Pajares, Matt Siervert, Vladimir Skokov, Raju Venugopalan, Douglas Wertepny, Urs Wiedemann and William Zajc, and the participants of the 2018 Benasque workshop on {\it Collectivity and correlations in high-energy hadron and nuclear collisions}  and the
{\it COST workshop on collectivity in small systems}, for collaborations and many fruitful discussions. We would like to specially thank Pedro Agostini, Guillaume Beuf, Alex Kovner and Michael Lublinsky for a longstanding collaboration.

NA is supported by Ministerio de Ciencia e Innovaci\'on of Spain under projects FPA2017-83814-P and Unidad de Excelencia Mar\'{\i}a de Maetzu under project MDM-2016-0692, by the European research Council under project ERC-2018-ADG-835105 YoctoLHC, by Xunta de Galicia under project ED431C 2017/07, by Conseller\'{\i}a de Educaci\'on, Universidade e Formaci\'on Profesional as Centro de Investigaci\'on do Sistema universitario de Galicia (ED431G 2019/05), and by FEDER. TA is supported by Grant No. 2018/31/D/ST2/00666 (SONA\-TA 14 - National Science Centre, Poland). This work has been performed in the framework of COST Action CA 15213 ``Theory of hot matter and relativistic heavy-ion collisions" (THOR), MSCA RISE 823947 ``Heavy ion collisions: collectivity and precision in saturation physics''  (HI\-EIC) and has received funding from the European Un\-ion's Horizon 2020 research and innovation programme under grant agreement No. 824093.

 \bibliographystyle{epj}
 \bibliography{mybib}
\end{document}